\documentclass[prd,preprint,tightenlines,floatfix,showpacs,preprintnumbers,nofootinbib,eqsecnum,superscriptaddress]{revtex4-1}

 \usepackage[dvips,final]{graphicx}
  \usepackage{amssymb}
   \usepackage{amsmath}
    \usepackage{amsfonts}
     \usepackage{epsfig}
      \usepackage{bm}

\def\lsim{\mathrel{\rlap{\lower4pt\hbox{\hskip1pt$\sim$}}
    \raise1pt\hbox{$<$}}}         
\def\gsim{\mathrel{\rlap{\lower4pt\hbox{\hskip1pt$\sim$}}
    \raise1pt\hbox{$>$}}}         

\begin{document}

\title{Exclusive diffractive photon bremsstrahlung at the LHC}

\author{Piotr Lebiedowicz}
\email{Piotr.Lebiedowicz@ifj.edu.pl}
\affiliation{Institute of Nuclear Physics PAN, PL-31-342 Cracow, Poland}

\author{Antoni Szczurek}
\email{Antoni.Szczurek@ifj.edu.pl}
\affiliation{Institute of Nuclear Physics PAN, PL-31-342 Cracow, Poland}
\affiliation{University of Rzesz\'ow, PL-35-959 Rzesz\'ow, Poland}

\begin{abstract}
We calculate differential distributions for the $p p \to p p \gamma$ reaction at 
the LHC energy $\sqrt{s} = 14$~TeV.
We consider diffractive classical bremsstrahlung mechanisms including effects
of non point-like nature of protons.
In addition, we take into account (vector meson)-pomeron, photon-pion
as well as photon-pomeron exchange processes 
for the first time in the literature.
Predictions for the total cross section and 
several observables related to these processes
e.g. differential distributions in pseudorapidities and transverse momenta
of photons or protons are shown and discussed.
The integrated diffractive bremsstrahlung cross section
($E_{\gamma}>100$~GeV) is only of the order of $\mu$b.
We try to identify regions of the phase space where one of the mechanisms dominates.
The classical bremsstrahlung dominates 
at large forward/backward photon pseudorapidities, 
close to the pseudorapidities of scattered protons.
In contrast, the photon-pomeron (pomeron-photon) mechanism
dominates at midrapidities but the related cross section is rather small. 
In comparison the virtual-omega rescattering mechanism contributes at smaller angles of photons 
(larger photon rapidities).
Photons in the forward/backward region can be measured 
by the Zero Degree Calorimeters (ZDCs) installed in experiments at the LHC
while the midrapidity photons are difficult to measure
(small cross section, small photon transverse momenta). 
Protons could be measured by ALFA detector (ATLAS) or TOTEM detector at CMS.
The exclusivity could be checked with the help of main central detectors.
\end{abstract}


\maketitle

\section{Introduction}

Exclusive diffractive photon bremsstrahlung mechanism at high energies 
was almost not studied in the literature.
Because at high energy the pomeron exchange is the driving mechanism of bremsstrahlung 
it is logical to call the mechanisms 
described by the diagrams shown in Fig.\ref{fig:bremsstrahlung_diagrams_abcd}
diffractive bremsstrahlung to distinguish from the low-energy
bremsstrahlung driven by meson exchanges
\footnote{The photon bremsstrahlung was intensively studied
in nucleon-nucleon collisions at low energies 
(see e.g. \cite{Nakayama,Scholten} and references therein).
There the dominant mechanisms are nucleon current (off-shell nucleon) 
and/or mesonic current (photon emitted from the middle of exchanged mesons)
contributions driven by meson exchanges.}.
The exclusive photon production mechanism
is similar to $p p \to p p \omega$ \cite{CLSS}
and $p p \to p p \pi^{0}$ \cite{LS13} processes.
As discussed in the past the dominant hadronic bremsstrahlung-type 
mechanism is the Drell-Hiida-Deck (DHD) mechanism \cite{Deck}
for diffractive production of $\pi N$ final states
(for a nice review we refer to \cite{Alberi_Goggi} and references therein).



The $p p \to p p\gamma$ process at high energies was discussed only recently 
\cite{Khoze} and it was  proposed to use the exclusive photon bremsstrahlung 
to measure or estimate elastic proton-proton cross section at the LHC.
Only approximate formulas for the classical bremsstrahlung were given there.
The participating particles were treated there as point-like particles.
No differential distributions for the exclusive bremsstrahlung have been discussed. 

The photons radiated off the initial and final state protons 
can be seen by the Zero Degree Calorimeters (ZDCs)
that are installed at about $140$ meters on each side of the interaction region.
They will measure very forward neutral particles
in the pseudorapidity region $|\eta| > 8.5$ at the CMS \cite{ZDC_CMS}
and the ATLAS ZDCs provide coverage of the region $|\eta| > 8.3$ \cite{ZDC_ATLAS}
\footnote{Recently, the exclusive $pp \to nn \pi^{+}\pi^{+}$ reaction has been studied in Ref.\cite{LS11}.
This reaction can be also measured with the help of the ZDC detectors.
Very large cross sections has been found which
is partially due to interference of a few mechanisms.
Presence of several interfering mechanisms precludes
extraction of the elastic $\pi^+ \pi^+$ scattering cross section.}.
The forward detectors beyond pseudorapidities of $|\eta| > 3$ 
provide an efficient veto against neutral particle backgrounds in 
the ZDCs from diffractive and non-diffractive events (see \cite{Khoze}).
Furthermore the proposed Forward Shower Counters (FSCs),
to detect and trigger on rapidity gaps in diffractive events,
would improve the measurements at the LHC significantly \cite{efficiency}.
In addition to a measurement of the elastic $pp$ cross section
the bremsstrahlung photons could allow for the evaluation of 
the total $pp$ cross section, luminosity 
and relative alignment of the ZDCs and of the Roman Pot detectors.





In this paper, we wish to present a first detailed studies of single photon
bremsstrahlung in the exclusive process $p p \to p p \gamma$.
We shall include classical bremsstrahlung diagrams 
as well as some new diagrams characteristic exclusively for
proton-proton scattering, not present e.g. in $e^+ e^-$ scattering.
We include diagrams which arise in the vector-dominance model as well as
photon-pion (pion-photon) and photon-pomeron (pomeron-photon)
exchange processes not discussed so far in the literature.
We shall try to identify the region of the phase space where one 
can expect a dominance of one of the processes
through detailed studies of several differential distributions.


\section{The amplitudes for the \boldmath $p p \to p p \gamma$ reaction}
\subsection{Diffractive classical bremsstrahlung mechanisms}

\begin{figure}[!ht]
a)\includegraphics[width=4.5cm]{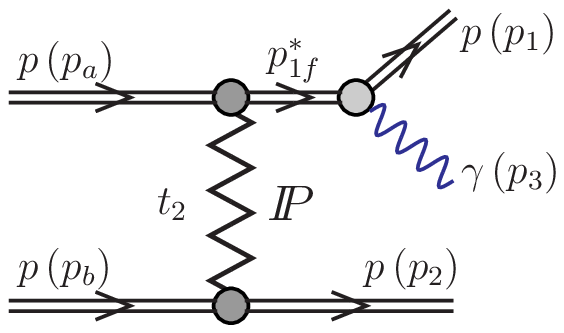}
c)\includegraphics[width=4cm]{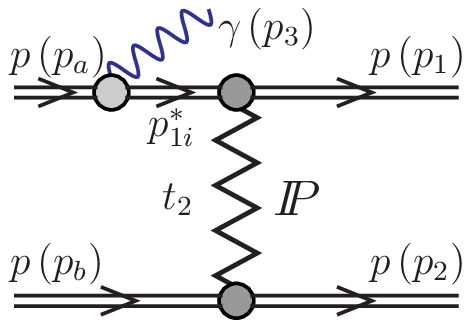}\\
b)\includegraphics[width=4.5cm]{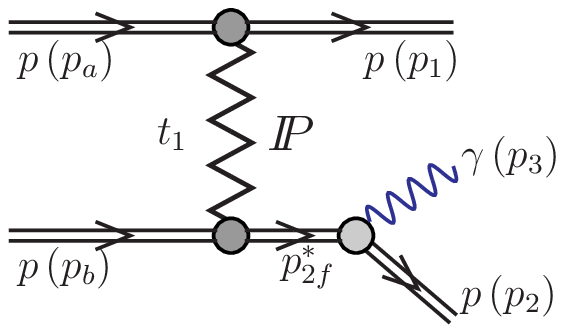}
d)\includegraphics[width=4cm]{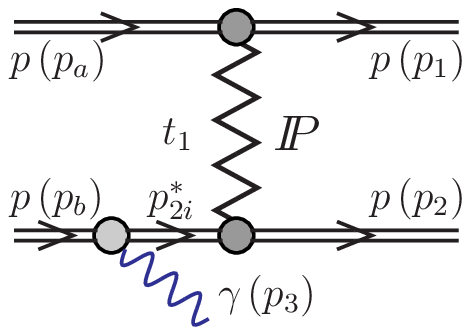}
   \caption{\label{fig:bremsstrahlung_diagrams_abcd}
\small Diagrams of the bremsstrahlung amplitudes driven by the pomeron exchange.
}
\end{figure}
The bremsstrahlung mechanisms for exclusive production of photons
discussed here are shown schematically in 
Fig.\ref{fig:bremsstrahlung_diagrams_abcd}.
In the case of $\gamma$ production 
the diagrams with intermediate nucleon resonances (see \cite{PDG}) 
should be negligible. 
The pronounced at low energy proton to $\Delta$ isobar transitions 
are suppressed in high energy regime. 

The Born amplitudes of diagrams shown in 
Fig.\ref{fig:bremsstrahlung_diagrams_abcd} can be written as
%
\begin{eqnarray}
{\cal M}^{(a)}_{\lambda_{a}\lambda_{b} \to \lambda_{1}\lambda_{2}\lambda_{3}} &=&
e\;
\bar{u}(p_{1},\lambda_{1}) 
\varepsilon^{*}\!\!\!\!\!/ \,(p_{3},\lambda_{3})
S_{N}(p_{1f}^{2})
\gamma^{\mu}
u(p_{a},\lambda_{a})\;
F_{\gamma N^{*}N}(p_{1f}^{2}) \;
F_{I\!\!P NN^{*}}(p_{1f}^{2}) \nonumber \\
&\times &
i s \,C_{I\!\!P}^{NN} \left( \frac{s}{s_{0}}\right)^{\alpha_{I\!\!P}(t_{2})-1} 
\exp\left(\frac{B_{I\!\!P}^{NN} t_{2}}{2}\right)\,\frac{1}{2s}\;
\bar{u}(p_{2},\lambda_{2}) 
\gamma_{\mu} 
u(p_{b},\lambda_{b})\,,
\label{brem_a}\\
{\cal M}^{(b)}_{\lambda_{a}\lambda_{b} \to \lambda_{1}\lambda_{2}\lambda_{3}} &=&
e\;
\bar{u}(p_{2},\lambda_{2}) 
\varepsilon^{*}\!\!\!\!\!/ \,(p_{3},\lambda_{3})
S_{N}(p_{2f}^{2}) 
\gamma^{\mu}
u(p_{b},\lambda_{b})\;
F_{\gamma N^{*}N}(p_{2f}^{2}) \; 
F_{I\!\!P NN^{*}}(p_{2f}^{2}) \nonumber \\
&\times &
i s \,C_{I\!\!P}^{NN} \left( \frac{s}{s_{0}}\right)^{\alpha_{I\!\!P}(t_{1})-1} 
\exp\left(\frac{B_{I\!\!P}^{NN} t_{1}}{2}\right)\,\frac{1}{2s}\;
\bar{u}(p_{1},\lambda_{1}) 
\gamma_{\mu} 
u(p_{a},\lambda_{a})\,,
\label{brem_b}\\
{\cal M}^{(c)}_{\lambda_{a}\lambda_{b} \to \lambda_{1}\lambda_{2}\lambda_{3}} &=&
e\;
\bar{u}(p_{1},\lambda_{1})
\gamma^{\mu}
S_{N}(p_{1i}^{2}) 
\varepsilon^{*}\!\!\!\!\!/ \,(p_{3},\lambda_{3})
u(p_{a},\lambda_{a})\;
F_{\gamma NN^{*}}(p_{1i}^{2}) \; 
F_{I\!\!P N^{*}N}(p_{1i}^{2}) \nonumber \\
&\times &
i s_{12} \,C_{I\!\!P}^{NN} \left( \frac{s_{12}}{s_{0}}\right)^{\alpha_{I\!\!P}(t_{2})-1} 
\exp\left(\frac{B_{I\!\!P}^{NN} t_{2}}{2}\right)\,\frac{1}{2s_{12}}\;
\bar{u}(p_{2},\lambda_{2}) 
\gamma_{\mu} 
u(p_{b},\lambda_{b})\,,
\label{brem_c}\\
{\cal M}^{(d)}_{\lambda_{a}\lambda_{b} \to \lambda_{1}\lambda_{2}\lambda_{3}} &=&
e\;
\bar{u}(p_{2},\lambda_{2})
\gamma^{\mu}
S_{N}(p_{2i}^{2}) 
\varepsilon^{*}\!\!\!\!\!/ \,(p_{3},\lambda_{3})
u(p_{b},\lambda_{b})\;
F_{\gamma NN^{*}}(p_{2i}^{2}) \; 
F_{I\!\!P N^{*}N}(p_{2i}^{2}) \nonumber \\
&\times &
i s_{12}\, C_{I\!\!P}^{NN} \left( \frac{s_{12}}{s_{0}}\right)^{\alpha_{I\!\!P}(t_{1})-1}
\exp\left(\frac{B_{I\!\!P}^{NN} t_{1}}{2}\right)\,\frac{1}{2s_{12}}\;
\bar{u}(p_{1},\lambda_{1}) 
\gamma_{\mu} 
u(p_{a},\lambda_{a})\,,
\label{brem_d}
\end{eqnarray}
where $u(p,\lambda)$, $\bar{u}(p',\lambda')=u^{\dagger}(p',\lambda')\gamma^{0}$
are the Dirac spinors (normalized as $\bar{u}(p') u(p) = 2 m_{p}$) of 
the initial and outgoing protons with the four-momentum $p$ and the helicities $\lambda$.
The factor $\frac{1}{2s}$ or $\frac{1}{2s_{12}}$ appear here
as a consequence of using spinors.
The four-momenta squared of virtual nucleons in the middle of diagrams are defined as
$t_{1,2}=q_{1,2}^{2}=(p_{a,b}-p_{1,2})^{2}$,
$p_{1i,2i}^{2}=(p_{a,b}-p_{3})^{2}$,
$p_{1f,2f}^{2}=(p_{1,2}+p_{3})^{2}$
and $s_{ij}=W_{ij}^{2}=(p_{i}+p_{j})^{2}$ are squared invariant masses of the $(i,j)$ system.
The propagators of the intermediate nucleons can be written as
\begin{eqnarray}
S_{N}(p^{2})= {\frac{i(p\!\!\!/ + m_{p})}{p^{2} - m_{p}^{2}}} \,,
\label{propagator_nucleon}
\end{eqnarray}
where $p\!\!\!/ = p_{\mu} \gamma^{\mu}$.
The polarization vectors of real photon
($\varepsilon^{*}\!\!\!\!\!/ \,(p_{3},\lambda_{3}) = 
\gamma^{\nu} \varepsilon_{\nu}^{*}(p_{3},\lambda_{3})$)
are defined in the proton-proton center-of-mass frame
%
\begin{eqnarray}
\varepsilon_{\nu}(p_{3},\pm 1) = \frac{1}{\sqrt{2}}
(0, i \sin\phi \mp \cos\theta\cos\phi,
-i \cos\phi \mp \cos\theta\sin\phi,
\pm \sin\theta)\,,
\label{vectors}
\end{eqnarray}
where $\theta$ is the polar angle and $\phi$ is the azimuthal angle 
of a emitted photon.
It is easy to check that they fulfill the relation
$\varepsilon^{\nu}(p,\lambda) \varepsilon^{*}_{\nu}(p,\lambda) =-1$ and
$p^{\nu} \varepsilon_{\nu}(p,\lambda) = 0$.

We use interaction parameters of Donnachie-Landshoff \cite{DL92} 
with $C_{I\!\!P}^{NN} = 21.7$~mb
and the pomeron trajectory $\alpha_{I\!\!P}(t)$ linear in $t$
%
\begin{eqnarray}
\alpha_{I\!\!P}(t) = \alpha_{I\!\!P}(0)+\alpha'_{I\!\!P}\,t,\quad
\alpha_{I\!\!P}(0) = 1.0808,\quad \alpha'_{I\!\!P} = 0.25 \; \mathrm{GeV}^{-2}\,.
\label{pomeron_trajectory}
\end{eqnarray}
%
The pomeron slope can be written as
\begin{eqnarray}
B(s) = B_{I\!\!P}^{NN} + 2 \alpha'_{I\!\!P} \ln \left( \frac{s}{s_{0}}\right)\,,
\label{slope}
\end{eqnarray}
where we use $s_{0} = 1$~GeV$^2$ and 
$B_{I\!\!P}^{NN} = 9$~GeV$^{-2}$ which approximately describes a running slope
for proton-proton elastic scattering.
Since in our calculations we include effective pomerons,
i.e. pomerons describing approximately nucleon-nucleon elastic scattering,
no explicit absorption corrections have to be included in addition.

In the bremsstrahlung processes discussed here the intermediate nucleons are off-mass shell. 
In our approach the off-shell effects related
to the non-point-like protons in the intermediate state are included by 
the following simple extra form factors
\begin{eqnarray}
F(p^{2}) = \frac{\Lambda_{N}^{4}}{(p^{2}-m_{p}^{2})^{2} + \Lambda_{N}^{4}}\,.
\label{extra_ff}
\end{eqnarray}
This form was used e.g. in Ref.\cite{OTL01} for $\omega$ photoproduction.
In general, the cut-off parameters in the form factors are not known
but could be fitted in the future to the (normalized) experimental data.
From our general experience in hadronic physics we expect $\Lambda_{N} \sim 1$~GeV.
We shall discuss how the uncertainties of the form
factors influence our final results.

We could ``improve'' the parametrization of the amplitudes
(\ref{brem_c}) and (\ref{brem_d}) to reproduce the high-energy Regge dependence
by the factors
$\left( s_{13} / m_{p}^{2} \right)^{\alpha_{N}(p^{2}_{1i})-\frac{1}{2}}$ and
$\left( s_{23} / m_{p}^{2} \right)^{\alpha_{N}(p^{2}_{2i})-\frac{1}{2}}$, respectively,
where the nucleon trajectory is
$\alpha_{N}(p^{2}_{1i,2i})=-0.3 + \alpha'_{N} \, p^{2}_{1i,2i}$, with $\alpha'_{N}=0.9$ GeV$^{-2}$.
We leave the problem of consistent nucleon reggezation in the context of
high-energy photon bremsstrahlung for future studies.

\subsection{Bremsstrahlung of \mbox{\boldmath $\omega$} mesons}

\begin{figure}[!ht]
e)\includegraphics[width=4cm]{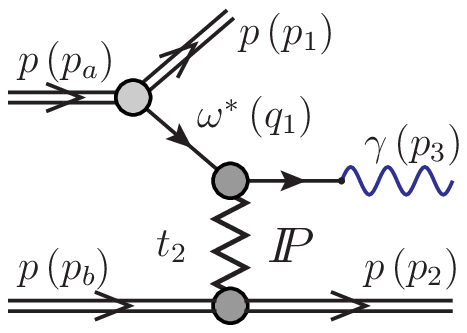}
f)\includegraphics[width=4cm]{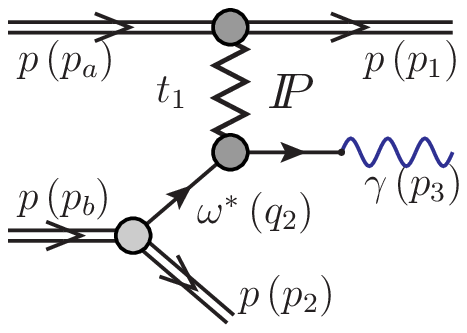}
   \caption{\label{fig:bremsstrahlung_diagrams_ef}
   \small Diagrams of the bremsstrahlung amplitudes with virtual-$\omega$ meson (reggeon)
and its transformation to final state photon.
}
\end{figure}

In Ref.~\cite{CLSS} we have discussed bremsstrahlung of $\omega$ mesons.
There one includes processes when $\omega$ meson emitted by an (anti)proton
interacts with the second (anti)proton.
The Born amplitudes for the interaction with emitted virtual-$\omega$ meson 
and its subsequent transformation to a photon,
shown in Fig.\ref{fig:bremsstrahlung_diagrams_ef},
are obtained as
\begin{eqnarray}
{\cal M}^{(e)\; \omega I\!\!P -exch.}_{\lambda_{a}\lambda_{b} \to \lambda_{1}\lambda_{2}\lambda_{3}} &=&
\bar{u}(p_{1},\lambda_{1}) 
\gamma^{\mu} 
u(p_{a},\lambda_{a}) 
S_{\mu \nu}(t_{1}) 
\varepsilon^{\nu*}(p_{3},\lambda_{3})\;
g_{\omega NN}\;  F_{\omega^{*}NN}(t_{1}) F_{I\!\!P \omega^{*}\omega}(t_{1})\, C_{\omega \to \gamma} \nonumber \\
&\times &
i s_{23} C_{I\!\!P}^{\omega N} \left( \frac{s_{23}}{s_{0}}\right)^{\alpha_{I\!\!P}(t_{2})-1}\;
\left( \frac{s_{13}}{s_{thr}}\right)^{\alpha_{\omega}(t_{1})-1}\;
\exp\left(\frac{B_{I\!\!P}^{\omega N} t_{2}}{2}\right)\; 
\delta_{\lambda_{2}\lambda_{b}}\,,
\label{brem_e}\\
%
{\cal M}^{(f)\; I\!\!P \omega -exch.}_{\lambda_{a}\lambda_{b} \to \lambda_{1}\lambda_{2}\lambda_{3}} &=&
\bar{u}(p_{2},\lambda_{2}) 
\gamma^{\mu} 
u(p_{b},\lambda_{b}) 
S_{\mu \nu}(t_{2})  
\varepsilon^{\nu*}(p_{3},\lambda_{3})\;
g_{\omega NN}\; F_{\omega^{*}NN}(t_{2}) F_{I\!\!P \omega^{*}\omega}(t_{2})\, C_{\omega \to \gamma}\; \nonumber \\
&\times &
i s_{13} C_{I\!\!P}^{\omega N} \left( \frac{s_{13}}{s_{0}}\right)^{\alpha_{I\!\!P}(t_{1})-1}\; 
\left( \frac{s_{23}}{s_{thr}}\right)^{\alpha_{\omega}(t_{2})-1}\;
\exp\left(\frac{B_{I\!\!P}^{\omega N} t_{1}}{2}\right)\;
\delta_{\lambda_{1}\lambda_{a}}\,,
\label{brem_f}
\end{eqnarray}
where $S_{\mu \nu}(t)$ is the propagator of the $\omega$ meson
\begin{eqnarray}
S_{\mu \nu}(t) = \frac{-g_{\mu \nu}+\frac{q_{\mu}q_{\nu}}{m_{\omega}^{2}}} {t - m_{\omega}^{2}}\,.
\label{propagator_omega}
\end{eqnarray}

In our calculation we assume
$C_{I\!\!P}^{\omega N} = C_{I\!\!P}^{\pi N} = 13.63$~mb \cite{DL92}
and the slope parameter 
$B^{\omega N}_{I\!\!P} = B^{\pi N}_{I\!\!P} = 5.5$~GeV$^{-2}$ (see e.g. \cite{CLSS,LS10}).
The amplitudes above, (\ref{brem_e}) and (\ref{brem_f}),
are corrected to reproduce the high-energy Regge dependence
by the Regge-like factors
$\left( s_{13}/s_{thr} \right)^{\alpha_{\omega}(t_1)-1}$ and
$\left( s_{23}/s_{thr} \right)^{\alpha_{\omega}(t_2)-1}$, respectively.
The $\omega$-reggeon trajectory is taken as
$\alpha_{\omega}(t) = 0.5 + 0.9\, t$ and
$s_{thr} = (m_{p}+m_{\omega})^2$.

Different values of the omega meson to nucleon coupling constant
have been used in the literature \cite{Bonn_potential}.
In our calculation we assume coupling constant
$g^2_{\omega N N}/4\pi = 10$. 
Similar value was used in \cite{Kaiser,NOHL07}. 
The transformation of $\omega$ meson to photon
is obtained within vector dominance model \cite{Szczurek_Uleshchenko}
and $C_{\omega \to \gamma} = \sqrt{\alpha_{em}/20.5} \simeq 0.02$, $\alpha_{em} = e^{2}/(4 \pi)$.
For completeness of this analysis we should include also
amplitudes for the interaction with emitted (virtual) $\rho$ meson.
Because of isospin, there is no mixing between the intermediate $\omega$ and $\rho$ mesons.
The transition of $\rho$ meson to photon 
is more probable $C_{\rho \to \gamma} = \sqrt{\alpha_{em}/2.54} \simeq 0.05$
whereas the coupling constant $g_{\rho N N}$ is small compared with $g_{\omega N N}$ --
in consequence the $\rho -I\!\!P$ contribution 
is comparable to the $\omega -I\!\!P$ contribution.

The off-shell form factors $F_{\omega^{*} NN}$ and $F_{I\!\!P \omega^{*} \omega}$
in (\ref{brem_e}) and (\ref{brem_f}) will be taken here in the following exponential form:
%
\begin{eqnarray}
F_{\omega^{*} NN}(t)&=&\exp\left(\frac{t-m_{\omega}^{2}}
{\Lambda_{\omega NN}^{2}}\right), \quad
F_{I\!\!P \omega^{*} \omega}(t)=\exp\left(\frac{t-m_{\omega}^{2}}
{\Lambda_{I\!\!P \omega \omega}^{2}}\right) \,,
\label{exp_form_factors}
\end{eqnarray}
where $\Lambda_{\omega NN} = \Lambda_{I\!\!P \omega \omega} = 1$~GeV.

\subsection{Pion cloud and \mbox{\boldmath $\gamma \pi^{0}$} and \mbox{\boldmath $\pi^{0} \gamma$} exchanges}

\begin{figure}[!ht]
g)\includegraphics[width=4.5cm]{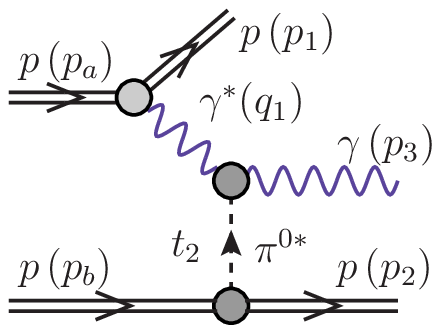}
h)\includegraphics[width=4.2cm]{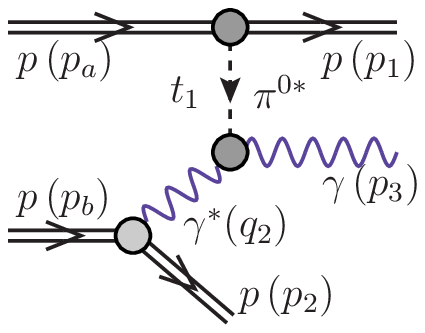}
   \caption{\label{fig:photon-pion_diagrams}
   \small  Diagrams with the $\gamma \pi^{0}$ and  
           $\pi^{0} \gamma$ exchanges in the $p p \to p p \gamma$ reaction.
}
\end{figure}

In our present analysis we include also 
$\gamma \pi^{0}$ and $\pi^{0} \gamma$ exchanges.
The underlying mechanisms are shown in
Fig.\ref{fig:photon-pion_diagrams}.
Such diagrams are dictated by the presence of pion cloud in the nucleon
(see e.g. \cite{HSS96}).

The amplitudes for the two new processes can be easily written as:
\begin{eqnarray}
{\cal M}^{(g)\; \gamma \pi^{0}-exch.}_{\lambda_{a}\lambda_{b} \to \lambda_{1}\lambda_{2}\lambda_{3}} &=&
e\,\bar{u}(p_{1},\lambda_{1}) \gamma^{\alpha} u(p_{a},\lambda_{a}) F_{1}(t_{1})\nonumber \\
&\times &
\dfrac{-g_{\alpha \beta}}{t_{1}}\,
F_{\gamma \pi \to \gamma}(t_1, t_2) \; 
\varepsilon^{\beta \mu \nu \lambda}\,
q_{1 \mu} \, p_{3 \nu}
\varepsilon_{\lambda}^{*}(p_{3},\lambda_{3})\nonumber \\
&\times &
g_{\pi^{0} NN} F_{\pi NN}(t_{2})\,
\dfrac{1}{t_{2}-m_{\pi}^{2}} \,
 \bar{u}(p_{2},\lambda_{2}) i \gamma_{5}
 u(p_{b},\lambda_{b})\,,
\label{gampi_amp_g}\\
%
{\cal M}^{(h)\; \pi^{0} \gamma -exch.}_{\lambda_{a}\lambda_{b} \to \lambda_{1}\lambda_{2}\lambda_{3}} &=&
g_{\pi^{0} NN} F_{\pi NN}(t_{1}) \,
\dfrac{1}{t_{1}-m_{\pi}^{2}} \,
\bar{u}(p_{1},\lambda_{1}) i \gamma_{5} u(p_{a},\lambda_{a})\nonumber \\
&\times &
\dfrac{-g_{\alpha \beta}}{t_{2}}\,
F_{\gamma \pi \to \gamma}(t_2, t_1) \; 
\varepsilon^{\beta \mu \nu \lambda}\,
q_{2 \mu} \, p_{3 \nu}
\varepsilon_{\lambda}^{*}(p_{3},\lambda_{3})\nonumber \\
&\times &
e\,\bar{u}(p_{2},\lambda_{2}) \gamma^{\alpha} u(p_{b},\lambda_{b}) F_{1}(t_{2})\,,
\label{pigam_amp_h}
\end{eqnarray}
where the $\gamma^{*} N N$-vertices are parametrized by the 
proton Dirac electromagnetic form factors 
%
%
%
\begin{eqnarray}
F_{1}(t)= \frac{4 m_{p}^{2}-2.79\,t}{(4 m_{p}^{2}-t)(1-t/m_{D}^{2})^{2}} \,,
\label{Fpomproton}
\end{eqnarray}
where $m_{p}$ is the proton mass and $m_{D}^{2} = 0.71$~GeV$^{2}$.
The coupling of the pion to the nucleon $g_{\pi NN}^{2}/4\pi = 13.5$ 
is relatively well known (see e.g.~\cite{ELT02}) 
and the corresponding hadronic form factor is taken in the exponential form
\begin{eqnarray}
F_{\pi NN}(t) = 
\exp \left( \frac{t - m_{\pi}^2}{\Lambda_{\pi NN}^2} \right)\,
\label{piNN_ff}
\end{eqnarray}
with $\Lambda_{\pi NN} = 1$~GeV.
For the central vertices involving off-shell particles
the $\gamma \pi^0$ form factors are taken in the
following factorized form
\begin{eqnarray}
F_{\gamma \pi \to \gamma}(t_1, t_2) &=& 
\dfrac{N_{c}}{12 \pi^{2} f_{\pi}}
\frac{m_{\rho}^2}{m_{\rho}^2 - t_1}
\exp \left( \frac{ t_2 - m_{\pi}^2 }{\Lambda_{\gamma \pi \to \gamma}^{2}}
\right)
\, 
\label{omega_pi_gamma_ff}
\end{eqnarray}
with the pion decay constant $f_{\pi} = 93$~MeV and $N_{c} = 3$.
The factor describing the virtual photon coupling
is taken as in the vector dominance model.
In practical calculations we take 
$\Lambda_{\gamma \pi \to \gamma} = 1$~GeV.


\subsection{Photon rescattering, \mbox{\boldmath $\gamma I\!\!P$} and \mbox{\boldmath $I\!\!P \gamma$} exchanges}

\begin{figure}[!ht]
i)\includegraphics[width=4.5cm]{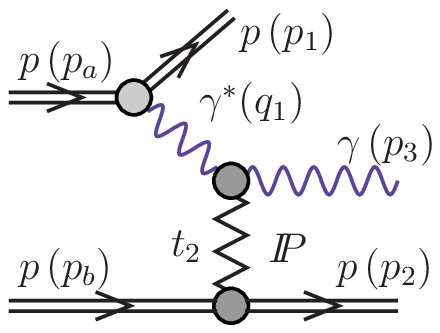}
j)\includegraphics[width=4.2cm]{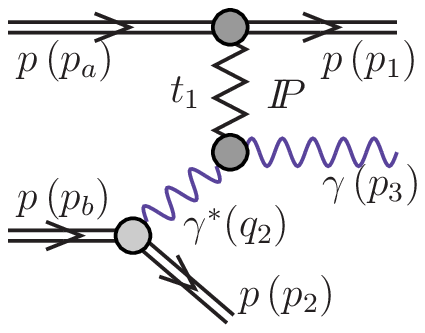}
   \caption{\label{fig:photon-pomeron_diagrams}
   \small  Diagrams with the $\gamma I\!\!P$ and  
$I\!\!P \gamma$ exchanges for the $p p \to p p \gamma$ reaction.
}
\end{figure}

At high energy there is still another type of diagrams (mechanisms) shown
in Fig.\ref{fig:photon-pomeron_diagrams}.
We shall call them in the following diagrams i) and j) for simplicity.
Here the intermediate photon couples to one of the protons through
electromagnetic form factors and interacts (at high energies) 
with the second proton exchanging pomeron (phenomenology) or 
gluonic ladder in the QCD language (see e.g.~\cite{SS07}).
This is a counterpart of the diagrams g) and h)
(see Fig.\ref{fig:photon-pion_diagrams}) relevant 
at lower $\gamma p$ subenergies.

The amplitude of the three-body process can be written in terms of 
the amplitude for elastic $\gamma p$ scattering.
For not too large $t$ the $\gamma p \to \gamma p$ amplitude
can be simply parametrized as
\begin{eqnarray}
{\cal M}_{\gamma p \to \gamma p}(s,t) \cong i s \sigma^{\gamma p}_{tot}(s)
\exp\left( \frac{B_{\gamma p}}{2} t \right) \,.
\label{gamp_gamp}
\end{eqnarray}
Such an amplitude gives, however, correct total cross section by construction. 
In the calculations presented in the Result section 
we shall use the simple Donnachie-Landshoff fit to the world data
on photon-proton total cross section \cite{DL92}
in which the pomeron and subleading reggeon exchanges
\footnote{In the reggeon contribution the $f_{2}$ exchange dominates over 
the $a_{2}$ exchange similarly as in the hadronic reactions, see \cite{DDLN}.}
have been included
%
%
\begin{eqnarray}
&&\sigma^{\gamma p}_{tot}(s) = C_{I\!\!P}^{\gamma p} s^{\alpha_{I\!\!P}(0)-1}
                             + C_{I\!\!R}^{\gamma p} s^{\alpha_{I\!\!R}(0)-1}\,, \nonumber \\
&& C_{I\!\!P}^{\gamma p} = 0.0677 \; \mathrm{mb}, \quad 
   C_{I\!\!R}^{\gamma p} = 0.129 \; \mathrm{mb}, \quad 
\alpha_{I\!\!P}(0) = 1.0808, \quad 
\alpha_{I\!\!R}(0) = 0.5475 \,.
\end{eqnarray}
%

In general, the slope parameter could be found by fitting to elastic 
$\gamma p$ scattering data which are, however, unknown and very difficult to measure. 
Since the incoming photon must first fluctuate to the $q \bar q$ state 
which interacts by the pomeron exchange with a proton
before forming the outgoing vector meson, 
it seems reasonable to use a hadronic slope for a first estimation. 
In practical calculations we shall use
$B_{\gamma p}(s) = B_{\pi p}(s)$ with its energy dependence, see Eq.(\ref{slope}).
 
Having fixed the elementary $\gamma p \to \gamma p$ amplitude 
we can proceed to our three-body photon rescattering amplitude. 
Limiting to large energies and small transverse momenta $t_{1}$ and $t_{2}$,
helicity conserving processes, the matrix element can be written as
\begin{eqnarray}
{\cal M}_{\lambda_{a} \lambda_{b} \to \lambda_{1} \lambda_{2} \lambda_{3}}
&\cong& \delta_{\lambda_2 \lambda_b}  
{\cal M}_{\gamma p \to \gamma p}(s_{23},t_{2})
\frac{e F_1(t_1)}{t_1} 
(p_{a}+p_{1})^{\mu}
\varepsilon^{*}_{\mu}(p_{3},\lambda_{3})
\delta_{\lambda_1 \lambda_a}
\nonumber \\
&+&       \delta_{\lambda_1 \lambda_a} 
{\cal M}_{\gamma p \to \gamma p}(s_{13},t_{1})
\frac{e F_1(t_2)}{t_2}
(p_{b}+p_{2})^{\mu}
\varepsilon^{*}_{\mu}(p_{3},\lambda_{3})
\delta_{\lambda_2 \lambda_b}
\,. 
\label{photon_pomeron}
\end{eqnarray}
%
Using $\vec{q}_{1,2 \perp} = - \vec{p}_{1,2 \perp}$ we have then
\begin{eqnarray}
{\cal M}_{\lambda_{a} \lambda_{b} \to \lambda_{1} \lambda_{2} \lambda_{3}}
&\cong& \delta_{\lambda_2 \lambda_b}  
{\cal M}_{\gamma p \to \gamma p}(s_{23},t_{2})
\frac{e F_1(t_1)}{t_1} 
\frac{2}{z_{1}} \frac{V^{*}(q_{1 \perp},\lambda_{3})}{\sqrt{1-z_{1}}}
\delta_{\lambda_1 \lambda_a}
\nonumber \\
&+&       \delta_{\lambda_1 \lambda_a} 
{\cal M}_{\gamma p \to \gamma p}(s_{13},t_{1})
\frac{e F_1(t_2)}{t_2}
\frac{2}{z_{2}} \frac{V^{*}(q_{2 \perp},\lambda_{3})}{\sqrt{1-z_{2}}}
\delta_{\lambda_2 \lambda_b}
\,,
\label{photon_pomeron_second_approach}
\end{eqnarray}
where the longitudinal momentum fractions of outgoing protons $z_{1,2}$ are
\begin{eqnarray}
z_1 \cong \frac{s_{23}}{s}, \quad z_2 \cong \frac{s_{13}}{s}, \quad z_{1}, z_{2} < 1 \,
\label{proton_fractions}
\end{eqnarray} 
and $V(q_{\perp}, \lambda_{3})$ can be calculated 
from $x,y$-components of momenta of participating protons
\begin{eqnarray}
V(q_{\perp}, \lambda_{3} = \pm 1) = \textbf{e}_{\mu}^{(\lambda_{3})} q_{\perp}^{\mu} 
= -\frac{1}{\sqrt{2}} \left( \lambda_{3} \, q_{x} + i q_{y} \right) \,.
\end{eqnarray} 
%

\subsection{Equivalent Photon Approximation}

The cross section for $\gamma \pi^{0}$ exchange and
$\gamma I\!\!P$ exchange mechanisms can be also calculated
in the Equivalent Photon Approximation (EPA).
In this approach the distribution of the photon can be written as
\begin{eqnarray}
\frac{d\sigma}{d y d p_{\perp}^2} &=&
z_1 f(z_1) \frac{d \sigma_{\gamma p \to \gamma p}}{dt_2}
\left(s_{23},t_2 \approx -p_{\perp}^2 \right)  \nonumber \\
&+& z_2 f(z_2) \frac{d \sigma_{\gamma p \to \gamma p}}{dt_1}
\left(s_{13},t_1 \approx -p_{\perp}^2 \right)\,,
\label{EPA}
\end{eqnarray}
where $f(z)$ is a photon flux in the proton; an explicit formula can 
be found e.g. in \cite{DZ89}.
The differential distribution of elastic scattering at high energies 
is parametrized as
\begin{eqnarray}
\frac{d \sigma_{\gamma p \to \gamma p}}{d t} \left(s,t \right)
&=& 
\frac{|{\cal M}_{\gamma p \to \gamma p}(s,t)|^2}{16 \pi s^{2}} \, .
\label{elastic_cross_section_gammap}
\end{eqnarray}
First energy and the longitudinal momentum of photon 
is calculated as a function of photon rapidity and transverse momentum $p_{\perp} = \sqrt{p_{x}^{2} + p_{y}^{2}}$
\begin{eqnarray}
&& E_{\gamma} = p_{\perp} \cosh{y}, \quad
   p_{z}= p_{\perp} \sinh{y}\,.
\end{eqnarray}
We get
\begin{eqnarray}
p_{z} =  \sqrt{E_{\gamma}^2 - p_{\perp}^2} \;\; \mathrm{for} \; y > 0, \;\; \mathrm{and} \;\;
p_{z} = -\sqrt{E_{\gamma}^2 - p_{\perp}^2} \;\; \mathrm{for} \; y < 0 \, .
\end{eqnarray}
Then energies in the photon-proton subsystems can be calculated approximately as
%
\begin{eqnarray}
s_{13} &\approx& (p_{a0} + E_{\gamma})^{2} - (p_{az} + p_{z})^{2} \,,\nonumber\\
s_{23} &\approx& (p_{b0} + E_{\gamma})^{2} - (p_{bz} + p_{z})^{2} \,.
\end{eqnarray}
The fractional energy, $z_{1}$ and $z_{2}$, losses of the protons
with four-momenta $p_{a}$ and $p_{b}$, respectively, can be obtained from Eq.(\ref{proton_fractions}).
%
\section{Results}

In the following section we shall show results of 
the differential distributions for the exclusive bremsstrahlung 
mechanisms discussed in the previous section.
The amplitudes for processes discussed in the sections above
are calculated numerically for each point in the phase space.
In calculating cross section of the three-body process
we perform integrations in
$\xi_1 = \log_{10}(p_{1\perp}/1\,\mathrm{GeV})$ 
and $\xi_2 = \log_{10}(p_{2\perp}/1\,\mathrm{GeV})$ 
instead in $p_{1\perp}$ and $p_{2\perp}$, 
in the photon (pseudo)rapidity $\eta_{\gamma}$
and the relative azimuthal angle between the outgoing protons
$\phi_{12} = \phi_{1} - \phi_{2}$.
%

\begin{figure}[!ht]
\includegraphics[width = 0.49\textwidth]{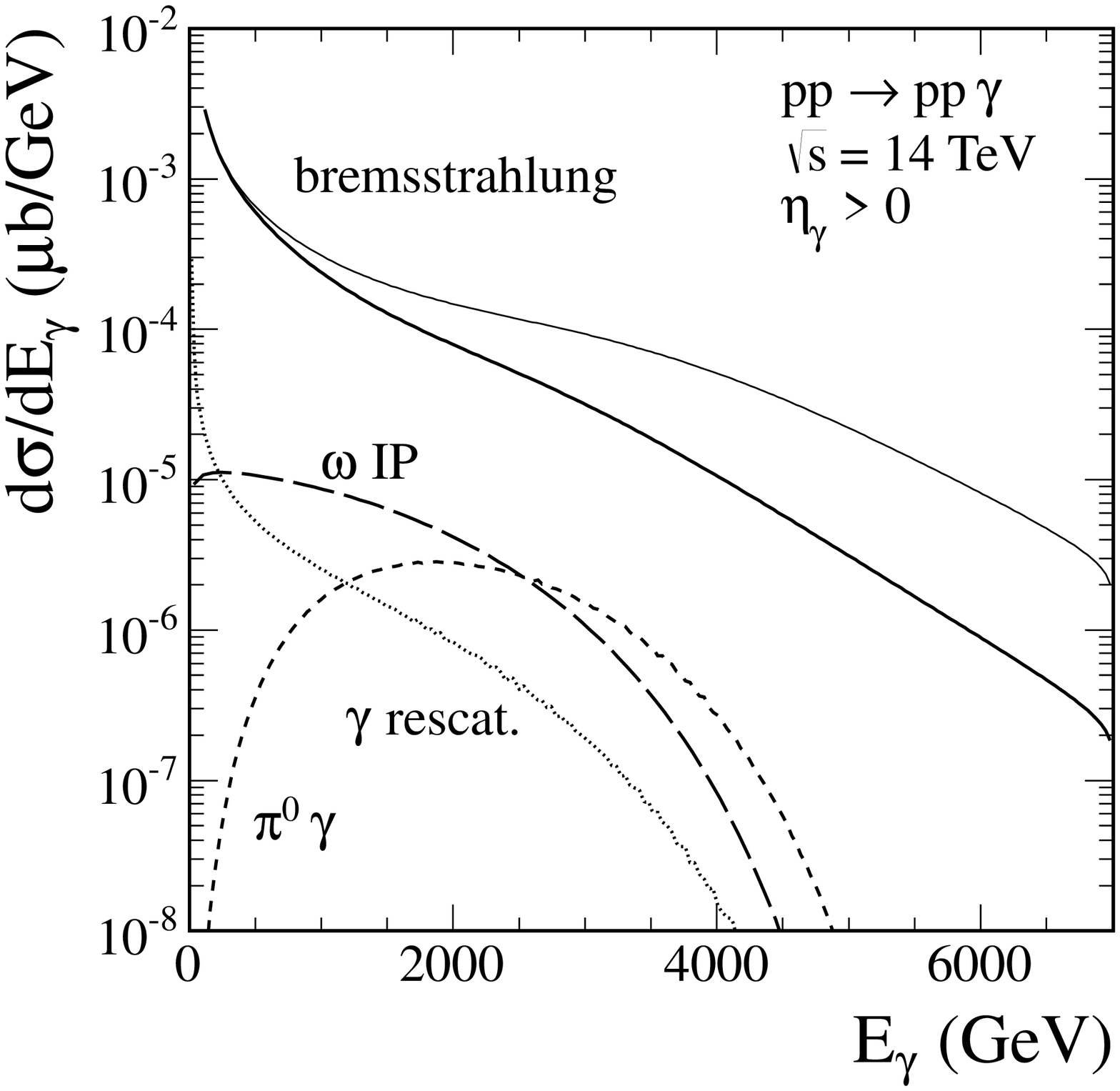}
\includegraphics[width = 0.49\textwidth]{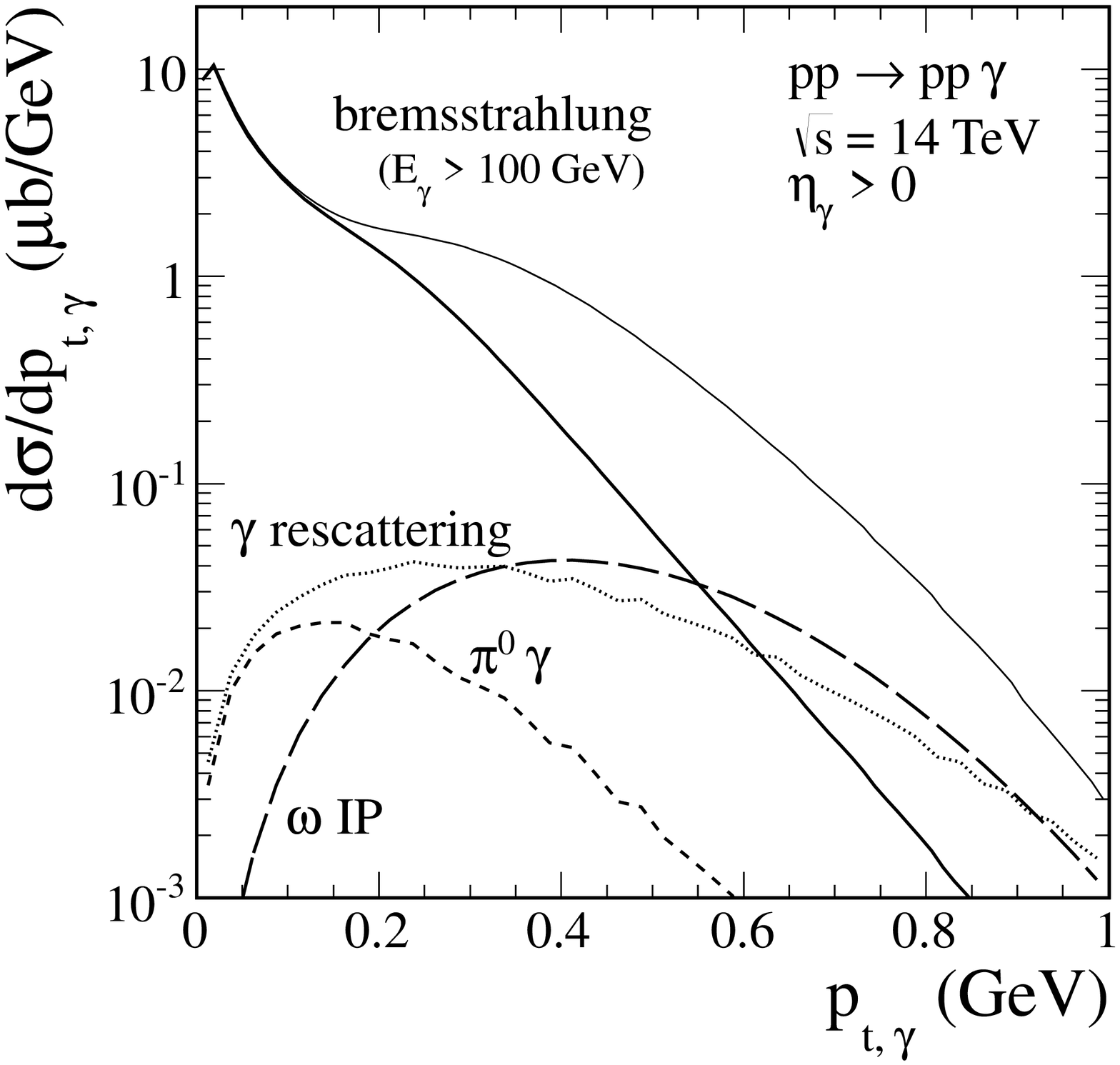}
  \caption{\label{fig:brem_dsig_dE3}
  \small
Energy spectrum of photons (left panel)
and distribution in transverse momentum of photons (right panel)
for all processes considered here at $\sqrt{s} = 14$~TeV and $\eta_{\gamma}>0$.
For classical bremsstrahlung we have imposed $E_{\gamma} > 100$~GeV
and used two values $\Lambda_{N} = 0.8, 1$~GeV of the proton off-shell form factors
(see the lower and upper solid line, respectively).
}
 \end{figure}

The photon energy spectrum drops relatively slowly
with photon energy as is shown in Fig.\ref{fig:brem_dsig_dE3} (left panel).
The ZDC detectors (at ATLAS or CMS) can measure only photons above some 
energy threshold (e.g. $E_{\gamma} > 50$~GeV).
In the calculation of classical bremsstrahlung
presented here we assume $E_{\gamma} > 100$~GeV as an example.
Corresponding distributions in the photon transverse momentum are 
shown in Fig.\ref{fig:brem_dsig_dE3} (right panel). The contribution of classical
bremsstrahlung is concentrated at very small transverse momenta
which is consistent with very small photon emission angle (large
pseudorapidity). The other distributions have rather similar shape
and vanish at $p_{\perp,\gamma} = 0$~GeV.
The exact shape may depend somewhat on the functional form
and values of cut-off parameters of off-shell
form factors taking into account the non-point-like nature of the vertices involved.
Here we have fixed the values of the corresponding form factors at typical hadronic scales.

\begin{figure}[!ht]
\includegraphics[width = 0.49\textwidth]{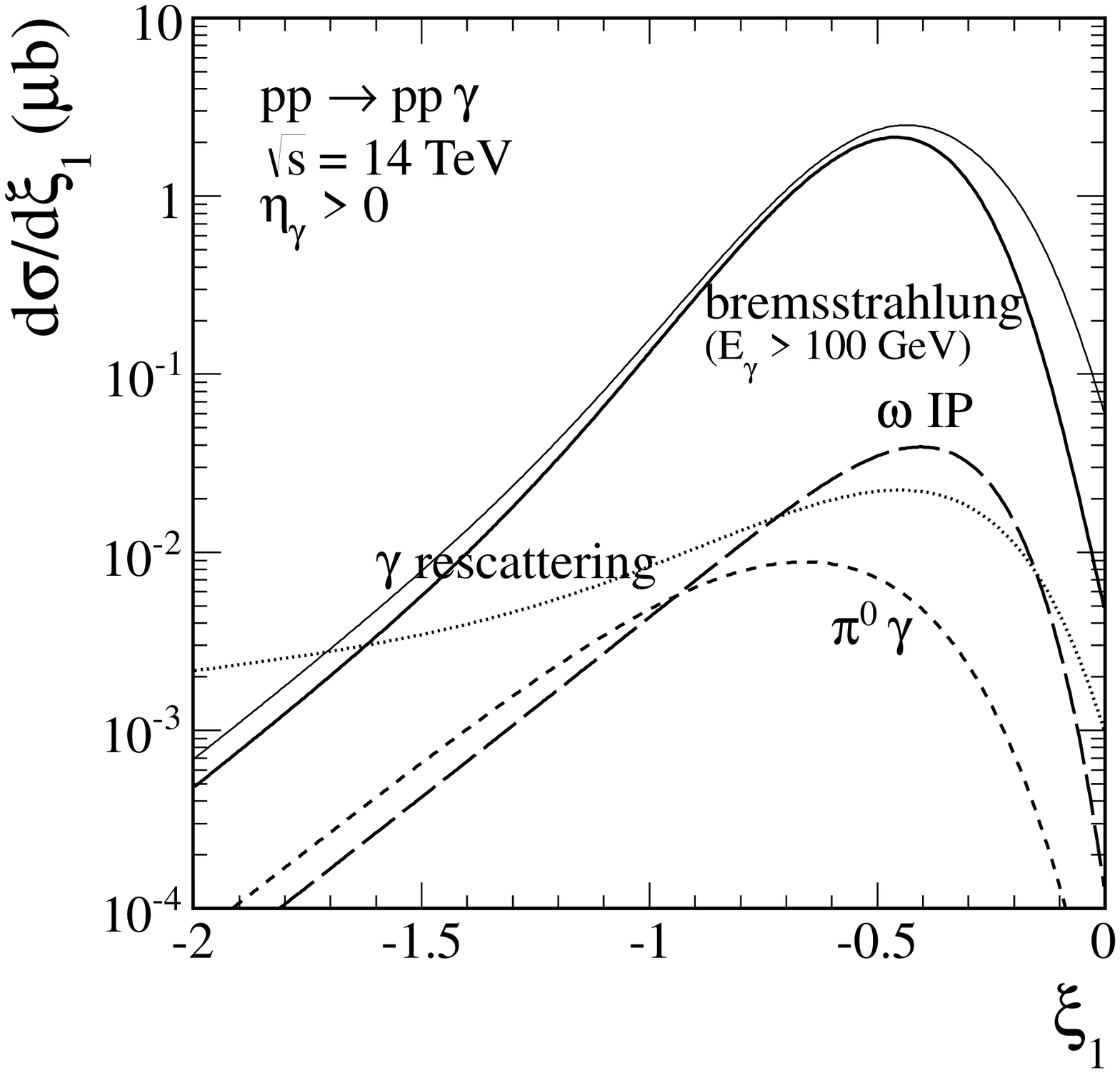}
\includegraphics[width = 0.49\textwidth]{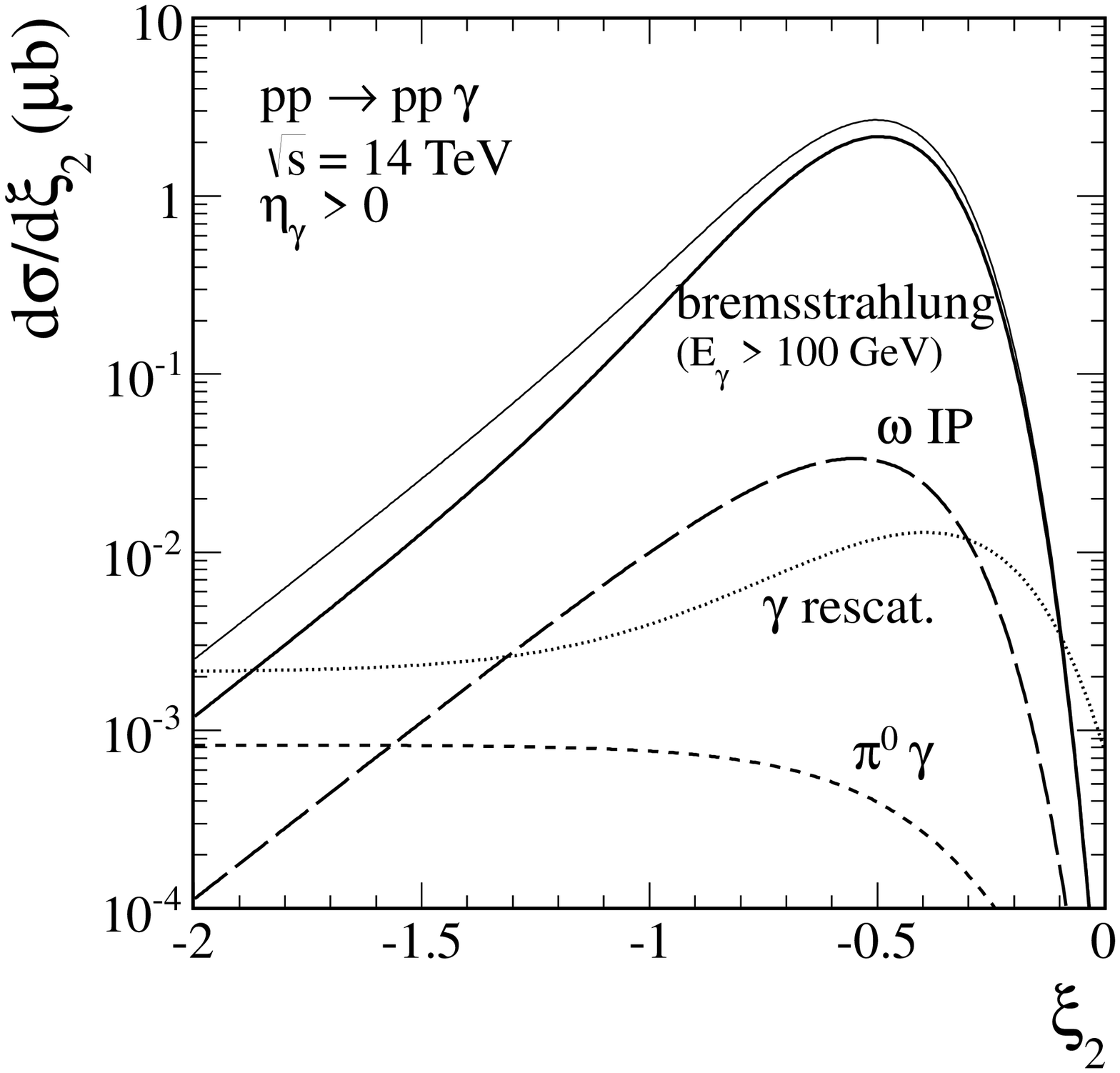}
  \caption{\label{fig:brem_dsig_dxi}
  \small
Distribution in $\xi_1 = \log_{10}(p_{1\perp}/1 \,\mathrm{GeV})$ (left panel)
and $\xi_2 = \log_{10}(p_{2\perp}/1 \,\mathrm{GeV})$ (right panel) at $\sqrt{s} = 14$~TeV and $\eta_{\gamma}>0$.
For classical bremsstrahlung $E_{\gamma} > 100$~GeV
and we have used two values of $\Lambda_{N} = 0.8, 1$~GeV (see the lower and upper
solid line, respectively) in the proton off-shell form factors.
}
 \end{figure}

In Fig.\ref{fig:brem_dsig_dxi} we show auxiliary distribution
in $\xi_1 = \log_{10}(p_{1\perp}/1\,\mathrm{GeV})$ (left panel) 
and $\xi_2 = \log_{10}(p_{2\perp}/1\,\mathrm{GeV})$ (right panel),
where $p_{1\perp}$ and $p_{2\perp}$ are outgoing proton transverse momenta.
For example $\xi = -1$ means proton transverse momenta $0.1$~GeV.
The biggest contribution for the classical bremsstrahlung process comes
from the region $\xi_i \approx - 0.5$ (i.e, $p_{i\perp} \approx 0.3$~GeV).
The distributions in $\xi_{1}$ or $\xi_{2}$ are different because we have limited
to the case of $\eta_{\gamma} > 0$ only.

\begin{figure}[!ht]
(a)\includegraphics[width = 7.cm]{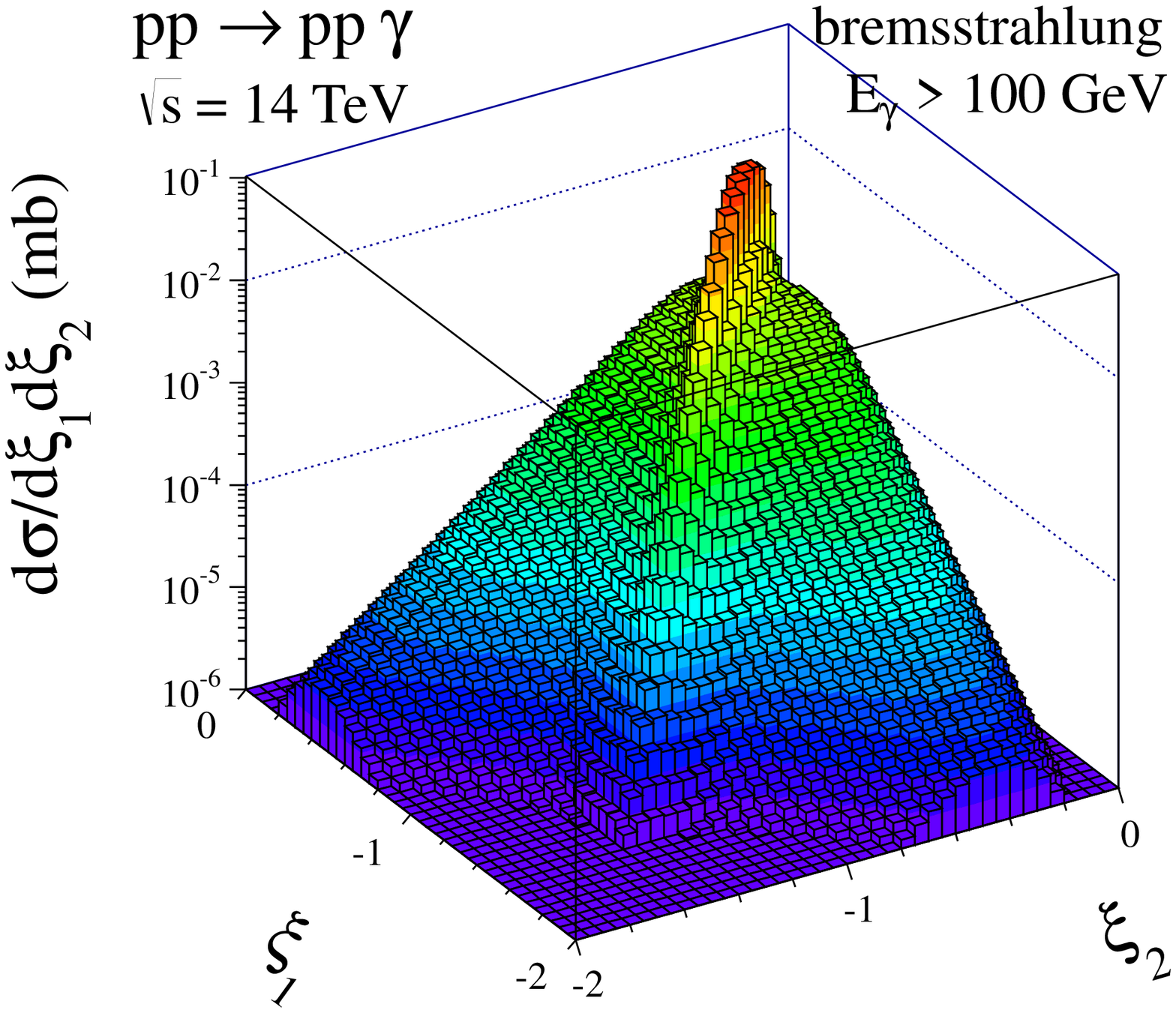}
(b)\includegraphics[width = 7.cm]{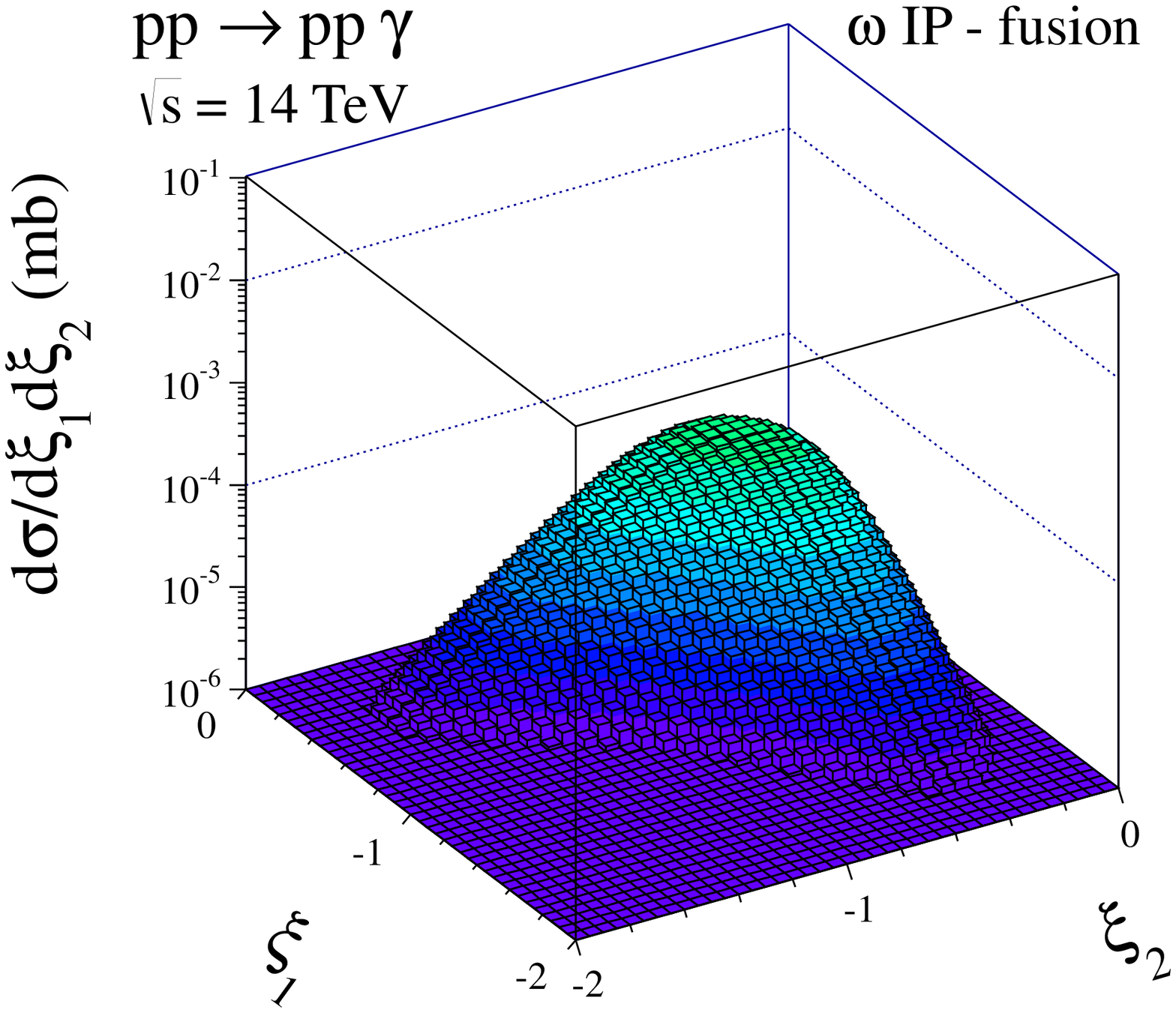}\\
(c)\includegraphics[width = 7.cm]{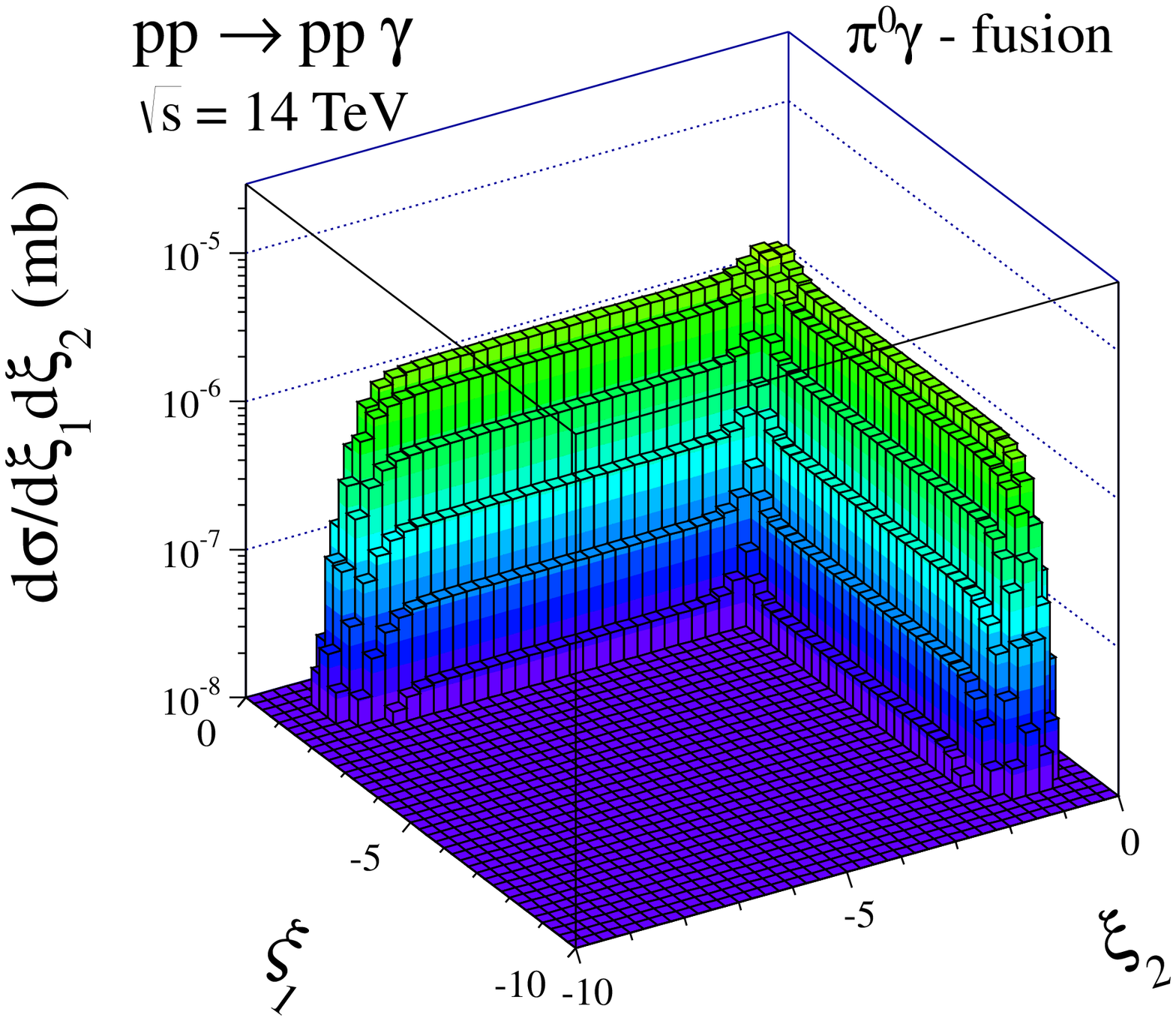}
(d)\includegraphics[width = 7.cm]{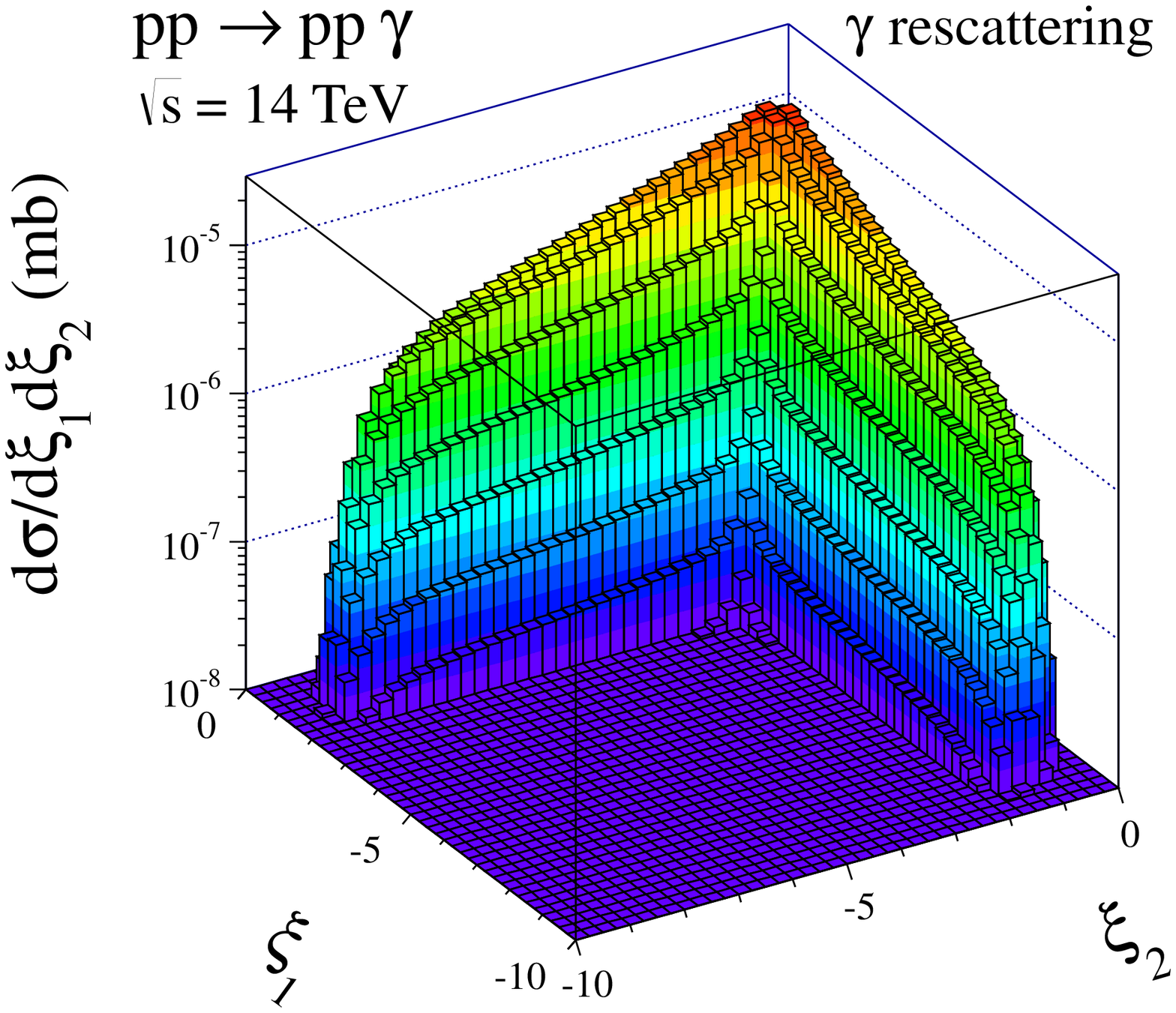}
  \caption{\label{fig:map_xi1xi2}
  \small
Distribution in 
$(\xi_1,\xi_2)$ = $(\log_{10}(p_{1\perp}/1\,\mathrm{GeV}), \log_{10}(p_{2\perp}/1\,\mathrm{GeV}))$
for the classical bremsstrahlung (a), the omega-rescattering (b), 
the pion-clould (c) and the photon-rescattering (d)
mechanisms at $\sqrt{s} = 14$~TeV.
For the classical bremsstrahlung we have imposed in addition $E_{\gamma} > 100$~GeV
and used $\Lambda_{N} = 1$~GeV in the proton off-shell form factors.
}
\end{figure}

In Fig.\ref{fig:map_xi1xi2} we show corresponding two-dimensional
distributions in $(\xi_1,\xi_2)$ in a full range of photon (pseudo)rapidity.
Quite different pattern can be seen for different mechanisms.
For the classical bremsstrahlung we observe an enhancement along the diagonal. 
This enhancement is a reminiscence of the elastic scattering for which $\xi_1 = \xi_2$.
Photon rescattering on the pion cloud (panel c) and 
photon rescattering with pomeron exchange (panel d)
are concentrated at small $\xi_{1}$ or $\xi_{2}$.

\begin{figure}[!ht]
\includegraphics[width = 0.49\textwidth]{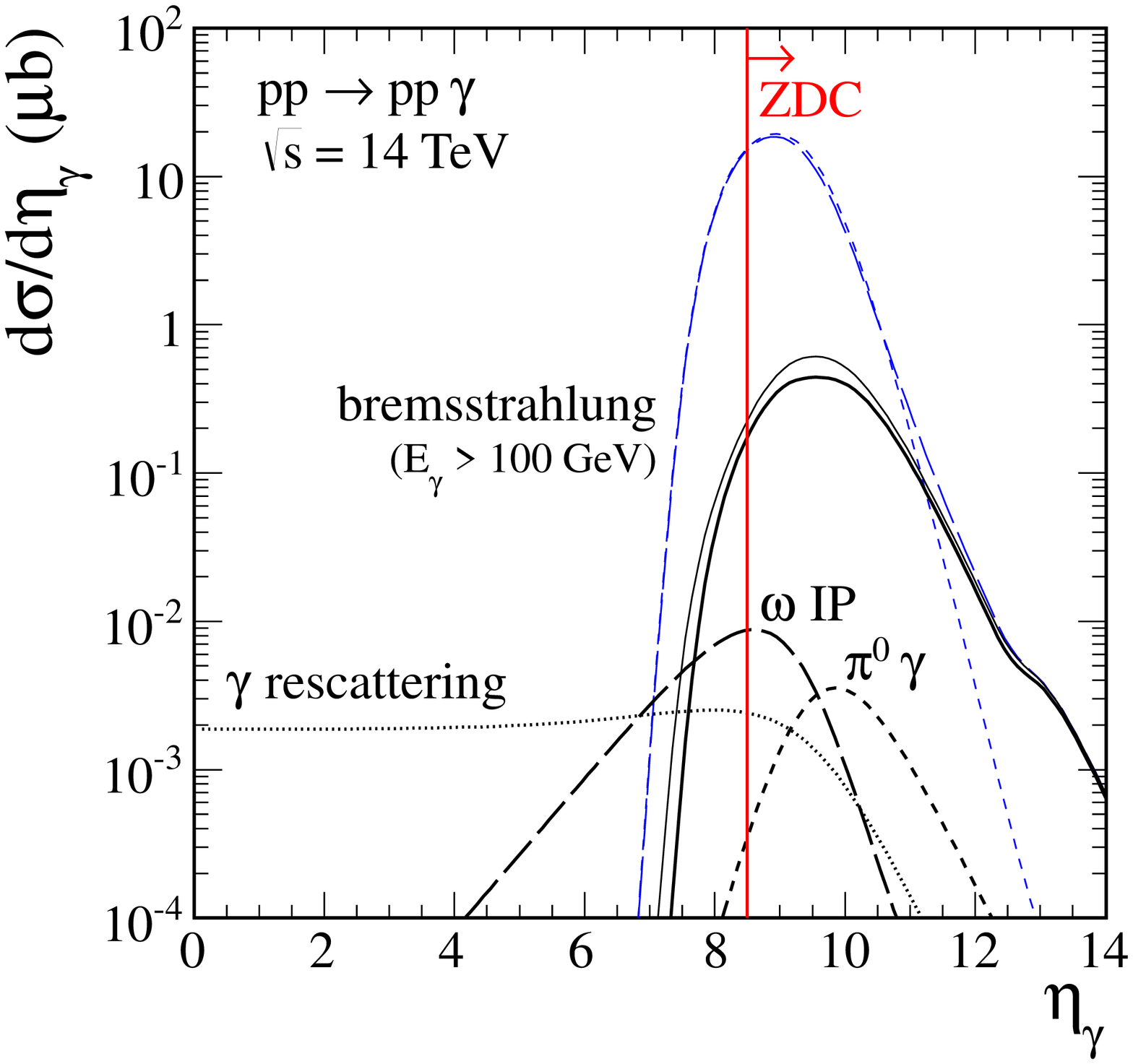}
\includegraphics[width = 0.49\textwidth]{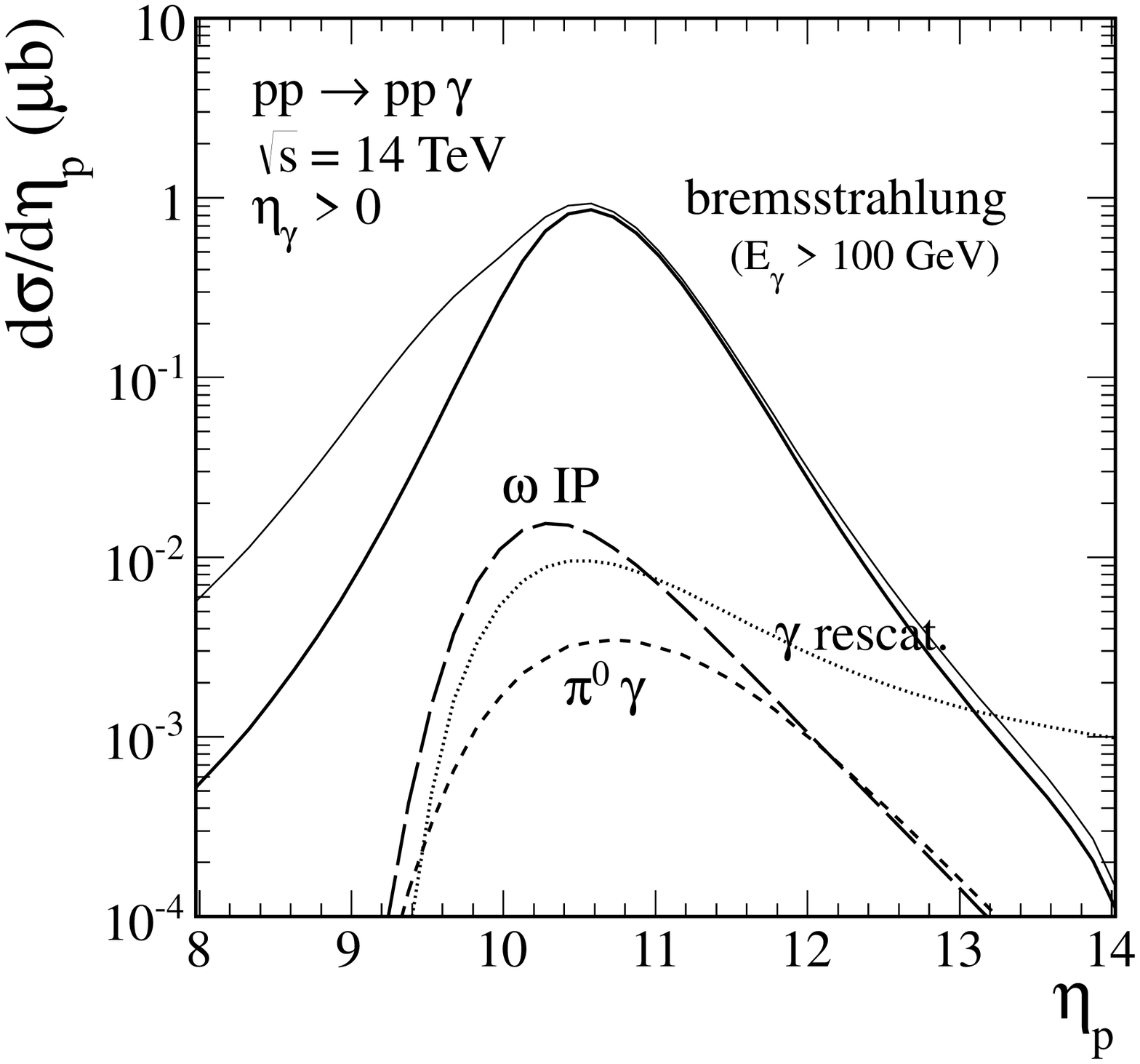}
  \caption{\label{fig:dsig_dy}
  \small
Distribution in (pseudo)rapidity of emitted photons (left panel) and 
in pseudorapidity of outgoing protons (right panel) at $\sqrt{s} = 14$~TeV.
For classical bremsstrahlung we have imposed $E_{\gamma} > 100$~GeV
and used two values of $\Lambda_{N} = 0.8, 1$~GeV in the proton off-shell form factors
(see the lower and upper solid line, respectively).
A large cancellation between the initial (\ref{brem_a}) and final state radiation (\ref{brem_c})
is shown (see the blue long-dashed and the blue short-dashed lines, respectively).
The lower pseudorapidity limit for the CMS ZDC detector is shown in addition by the vertical line.
}
\end{figure}

Photon (pseudo)rapidity distribution is particularly interesting.
In Fig.\ref{fig:dsig_dy} we show both distribution for photons $\eta_{\gamma}$ (left panel)
and corresponding distribution for outgoing protons $\eta_{p}$ (right panel)
for all processes considered in the present paper. 
In this variable both protons and photons 
are localized in a similar region of pseudorapidities (or equivalently polar angles).
The classical bremsstrahlung clearly gives the largest contribution.
It is also concentrated at very large $\eta{\gamma}$ i.e. 
in the region where ZDC detectors can be used.
We observe a large cancellation between the corresponding terms in the amplitude
(\ref{brem_a}) and (\ref{brem_c}) (see left panel) or 
(\ref{brem_b}) and (\ref{brem_d}).
The photon rescattering process clearly dominates in the region of $\eta{\gamma}$.
The cross section for this process is rather small.
Clearly an experimental measurement there would be a challenge.

\begin{figure}[!ht]
\includegraphics[width = 0.49\textwidth]{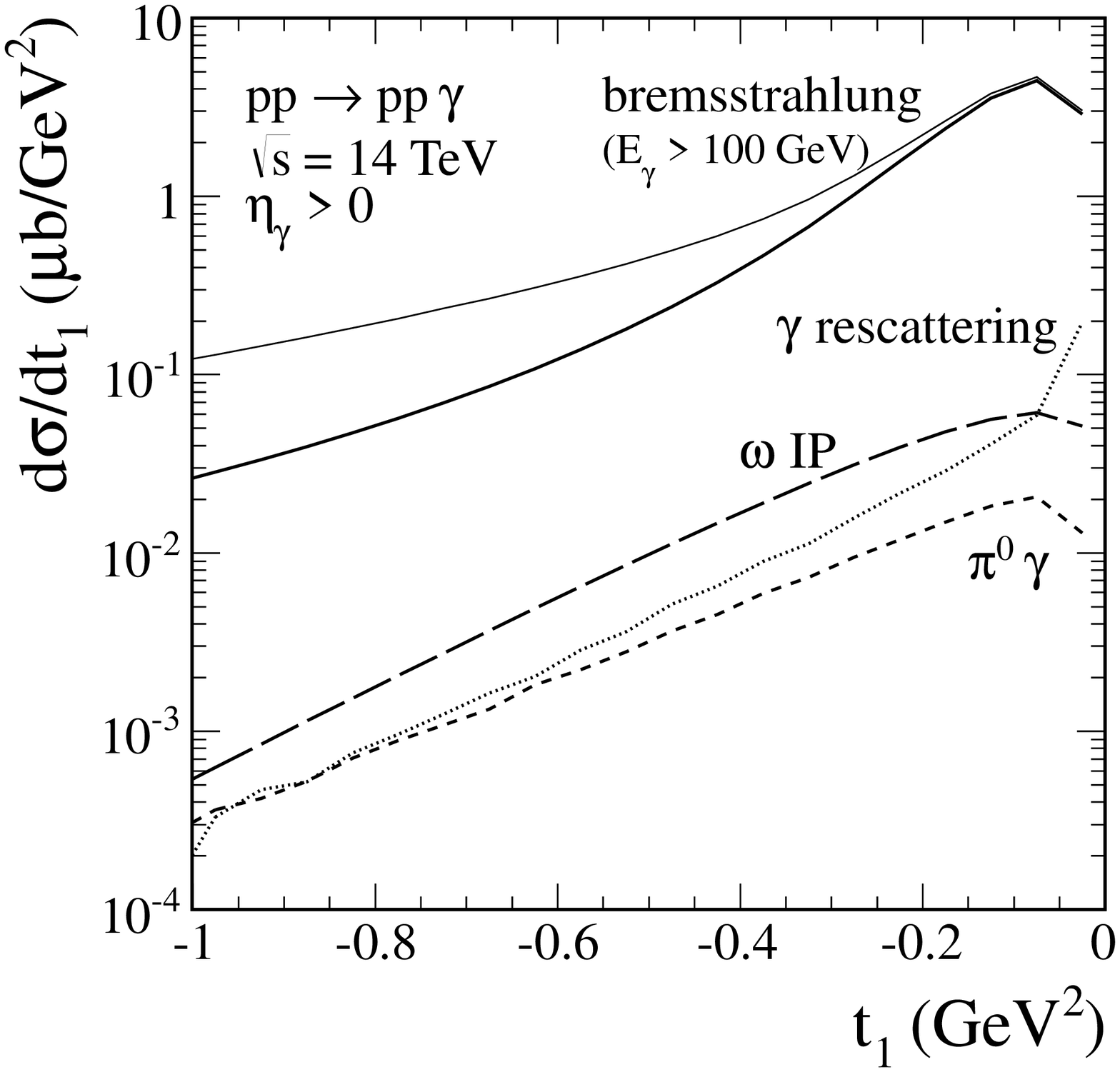}
\includegraphics[width = 0.49\textwidth]{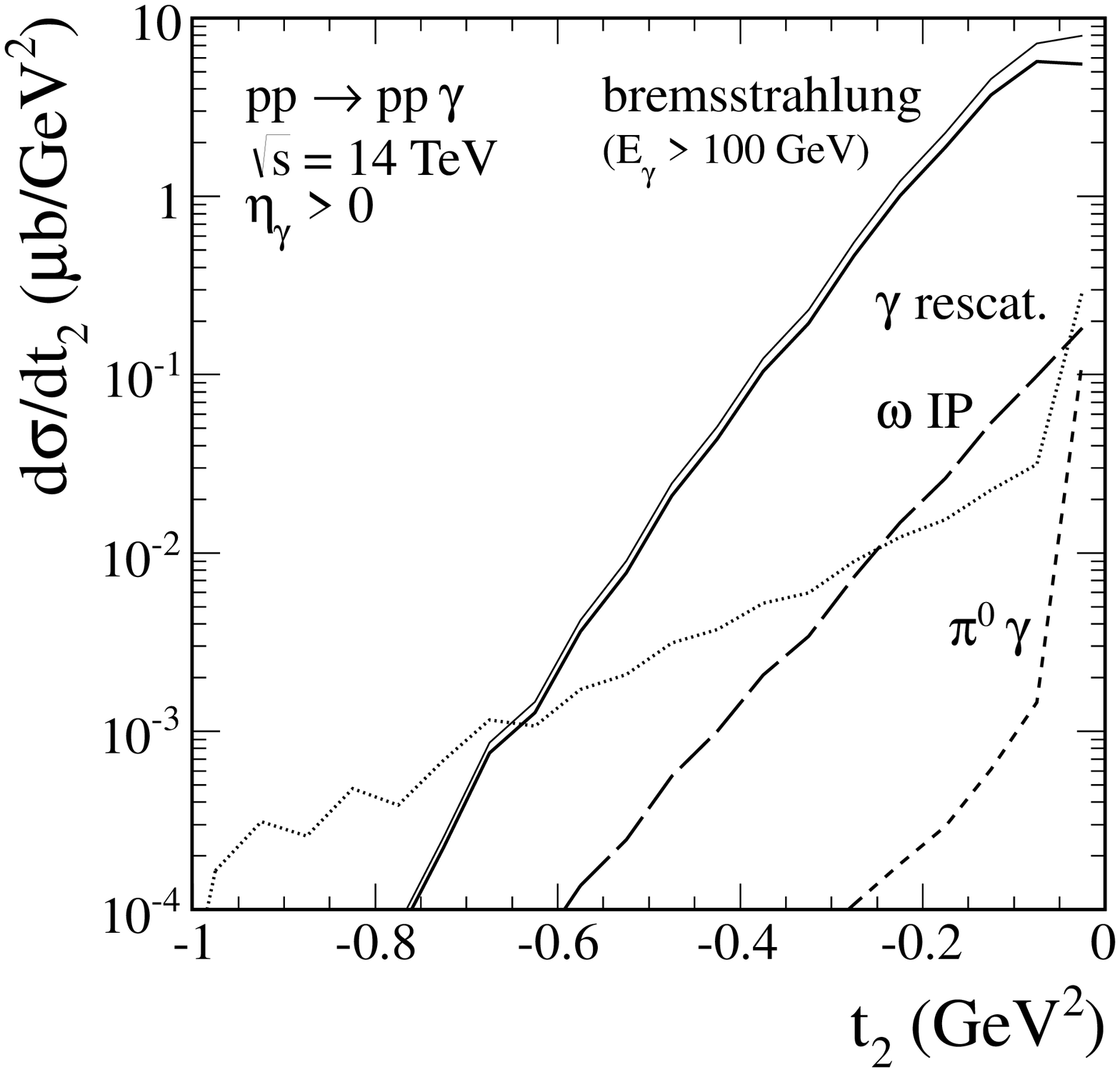}
  \caption{\label{fig:brem_dsig_dt}
  \small
Distribution in four-momentum squared $t_{1}$ (left panel) and $t_{2}$ (right panel)
at $\sqrt{s} = 14$~TeV and $\eta_{\gamma}>0$.
For classical bremsstrahlung we have imposed $E_{\gamma} > 100$~GeV
and used two values of $\Lambda_{N} = 0.8, 1$~GeV in the proton off-shell form factors
(see the lower and upper solid line, respectively).
}
 \end{figure}

In a first experimental trial one could measure only photons and perform
a check for rapidity gap in the midrapidity region. If protons are
measured in addition, one could analyze also some new observables related to protons.
In Fig.\ref{fig:brem_dsig_dt} we show distribution in 
the four-momentum transfer squared between initial and final protons.
One can observe a change of slope of the $t$ distribution
which is caused by the bremsstrahlung of photons.
In our simplified model we have assumed a constant (in $t_{1}$ and $t_{2}$) energy-dependent slope.

\begin{figure}[!ht]
(a)\includegraphics[width = 7.cm]{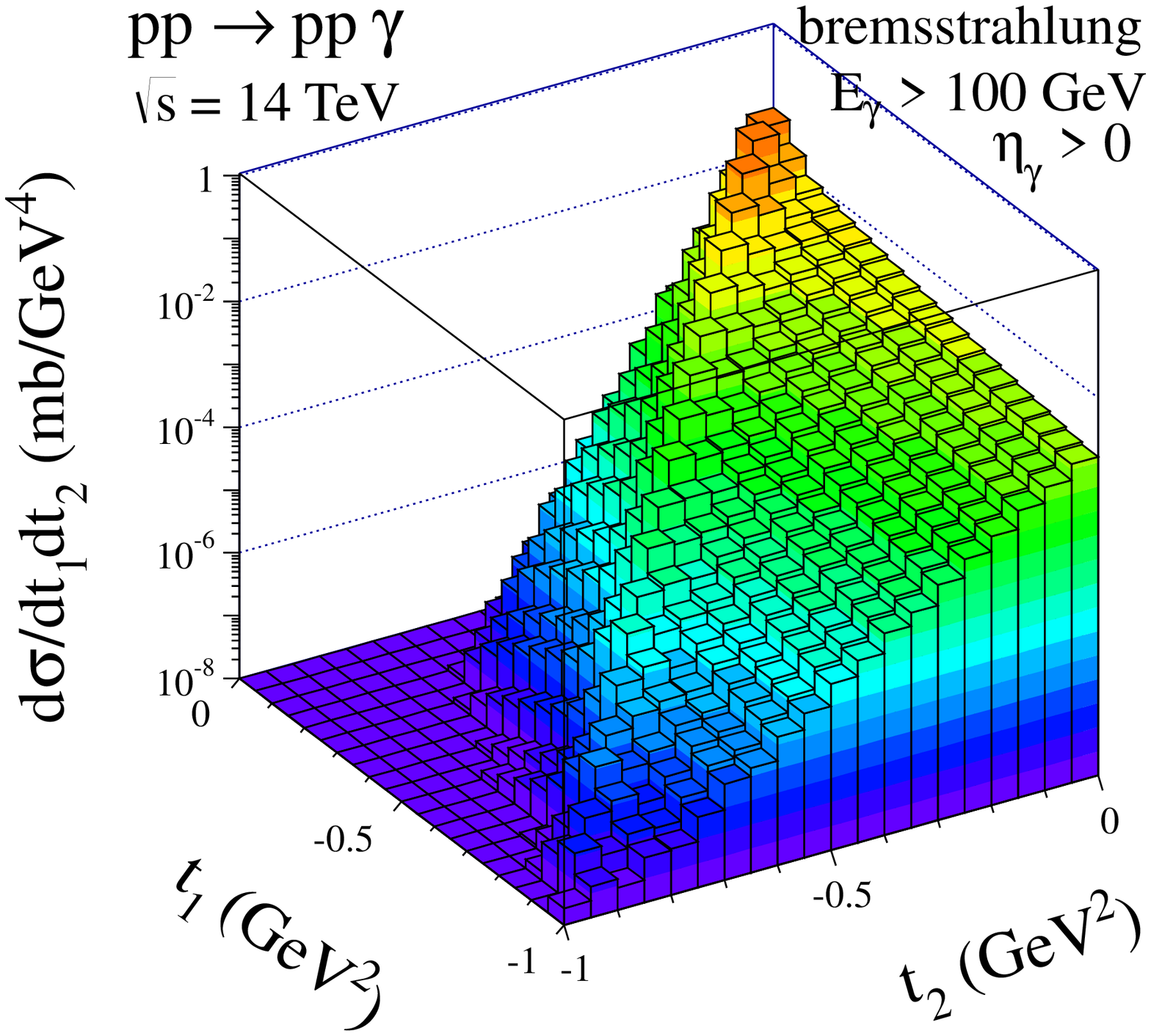}
(b)\includegraphics[width = 7.cm]{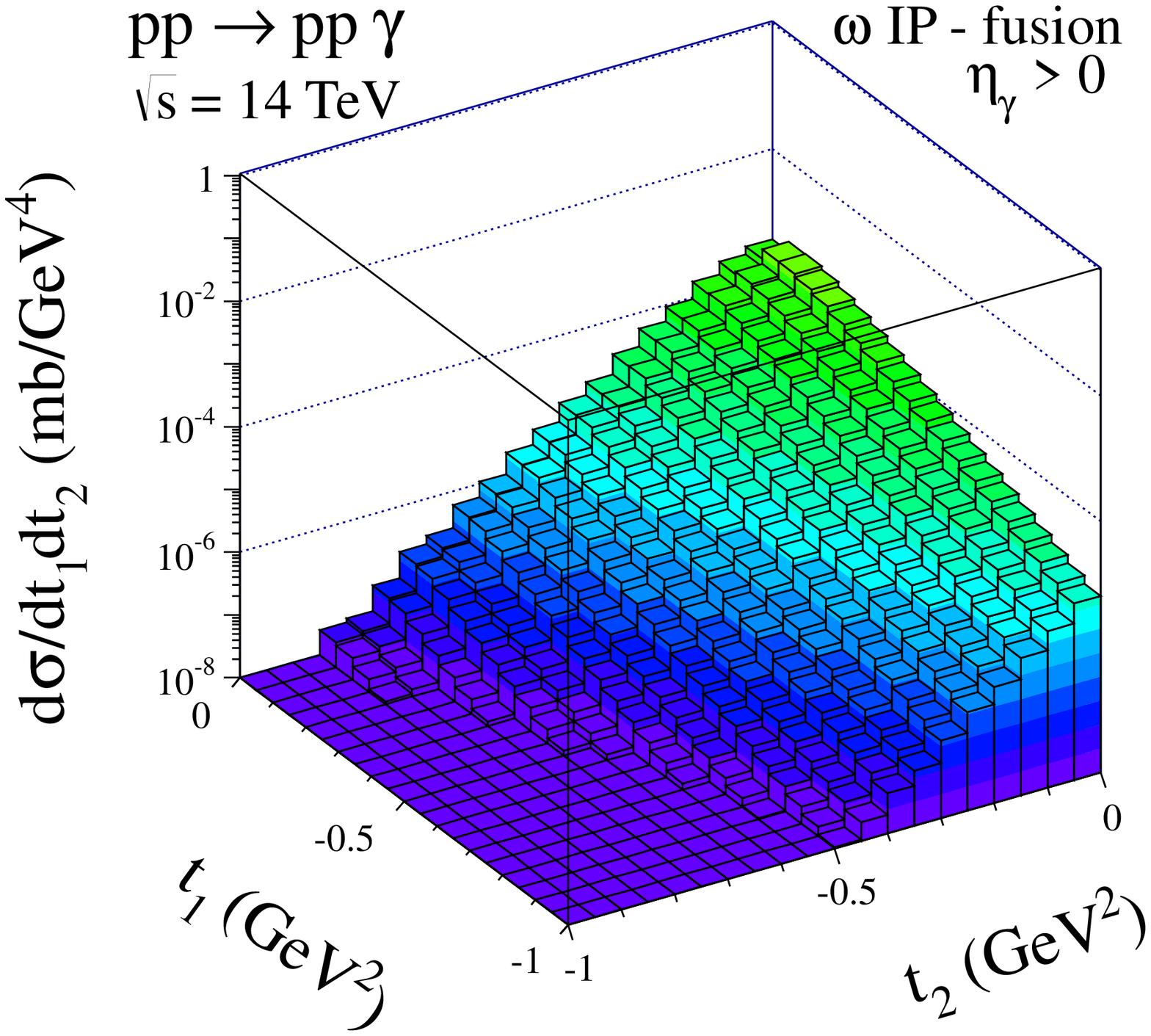}\\
(c)\includegraphics[width = 7.cm]{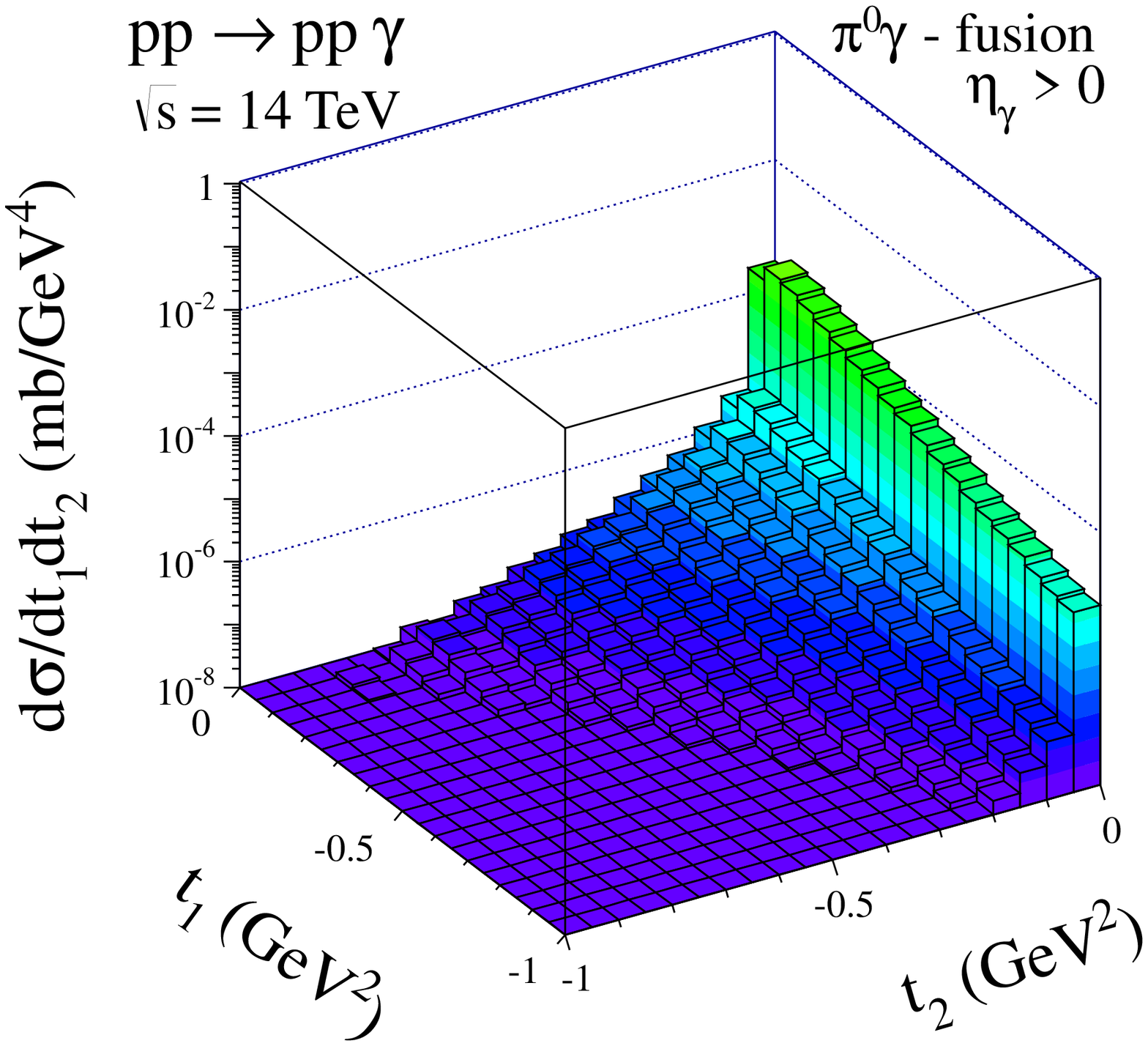}
(d)\includegraphics[width = 7.cm]{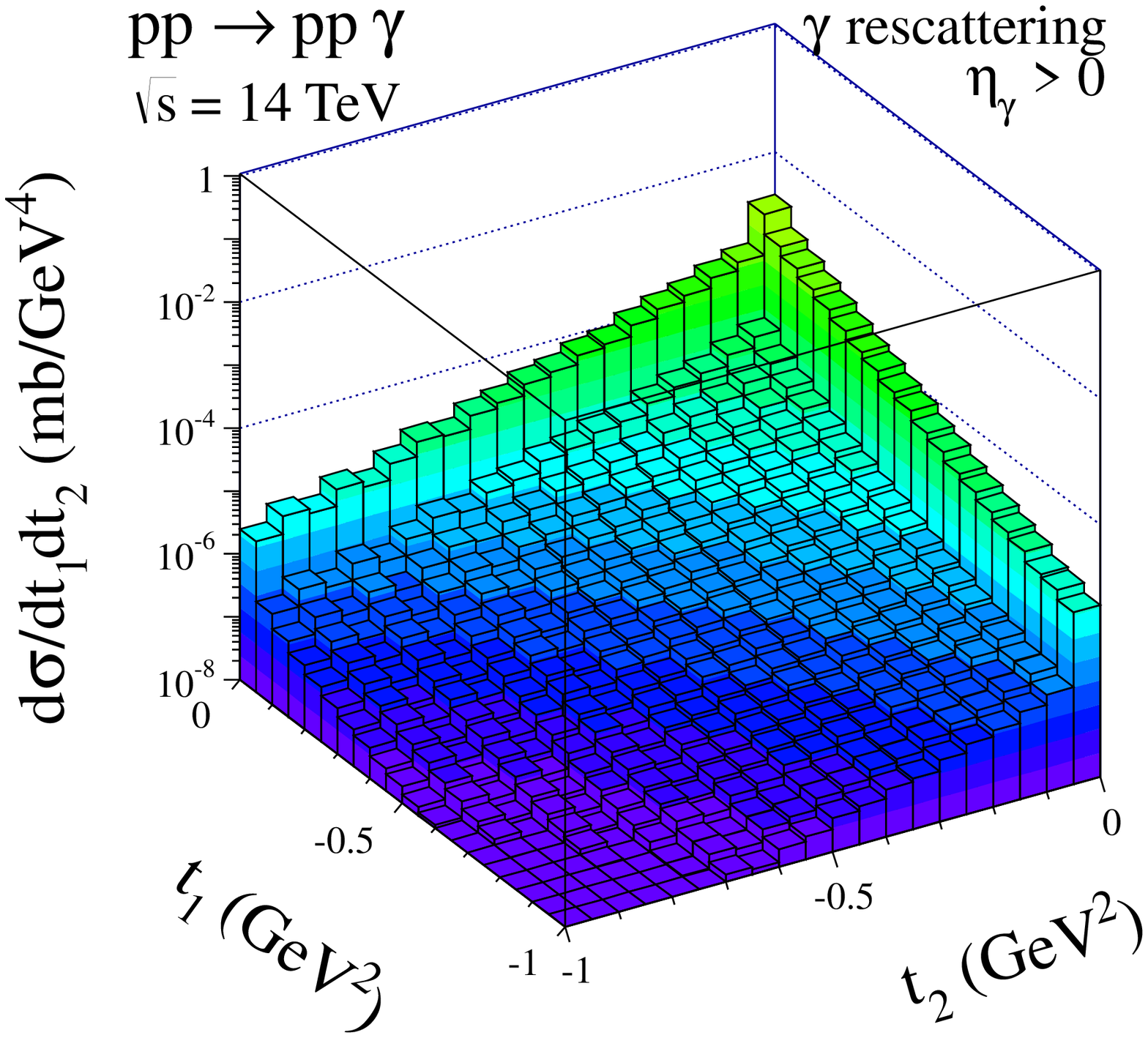}
  \caption{\label{fig:map_t1t2}
  \small
Distribution in ($t_1, t_2$) at $\sqrt{s} = 14$~TeV and $\eta_{\gamma}>0$
for the classical bremsstrahlung (a) ($E_{\gamma} > 100$~GeV and $\Lambda_{N} = 1$~GeV),
the omega rescattering (b), the pion cloud (c) and the photon rescattering (d) contributions.
}
 \end{figure}

In Fig.\ref{fig:map_t1t2} we show distribution in two-dimensional
space $(t_1,t_2)$. 
For the classical bremsstrahlung (left panel) one can
observe a ridge when $t_{1} \simeq t_{2}$
which is reminiscence of elastic scattering.
The distributions discussed here could in principle be obtained with 
the TOTEM detector at CMS to supplement the ZDC detector for the measurement of photons.

\begin{figure}[!ht]
\includegraphics[width = 0.49\textwidth]{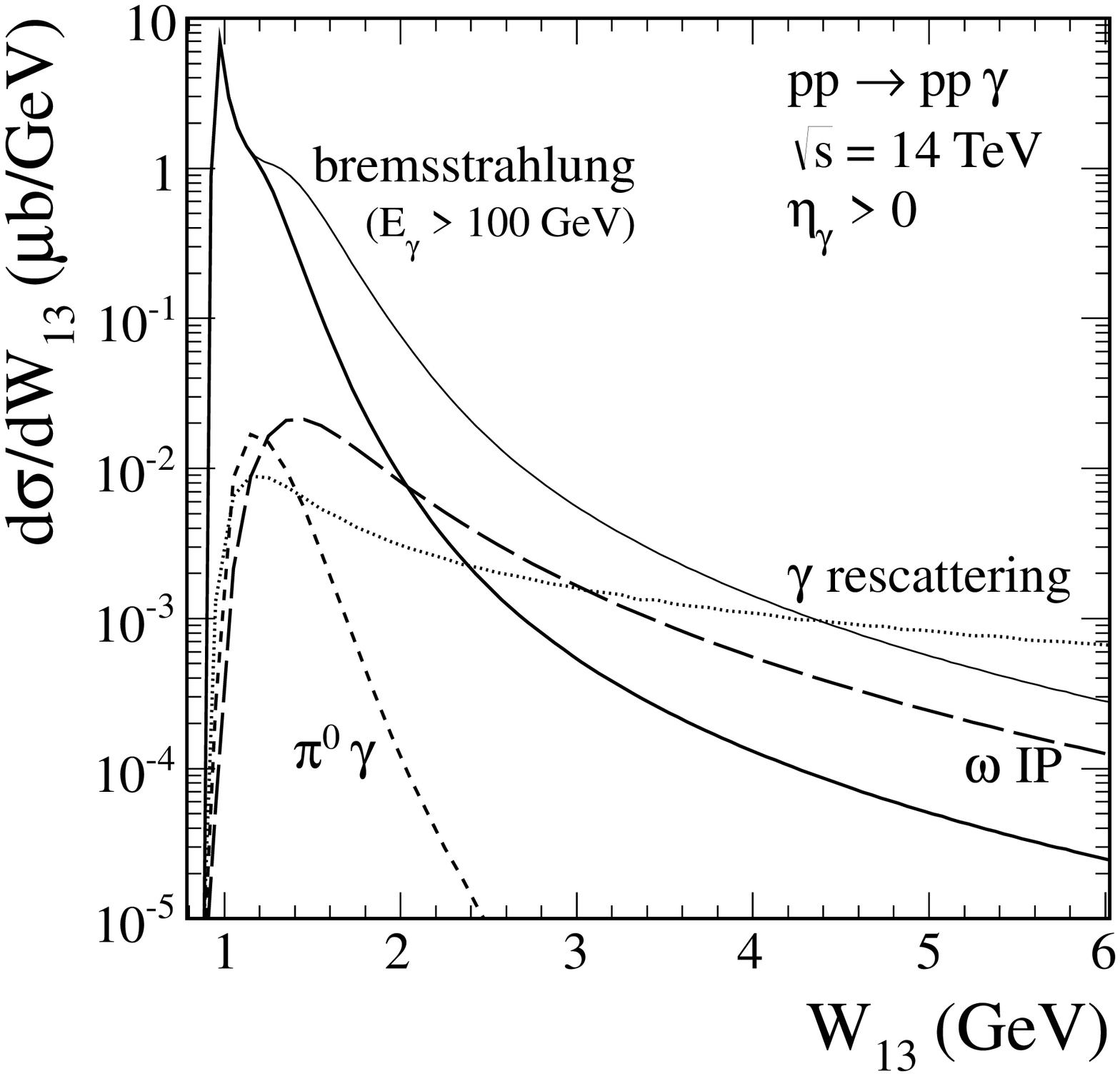}
  \caption{\label{fig:dsig_dw13}
  \small
Distribution in $\gamma p$ subsystem energy $W_{13}$ for all processes
considered here at $\sqrt{s} = 14$~TeV and $\eta_{\gamma}>0$.
For classical bremsstrahlung we have imposed $E_{\gamma} > 100$~GeV
and used two values of $\Lambda_{N} = 0.8, 1$~GeV of the proton off-shell form factors
(see the lower and upper solid line, respectively).
}
\end{figure}

In Fig.\ref{fig:dsig_dw13} we compare distribution 
in photon-(forward proton) subsystem energy 
for all processes considered in the present paper.
The discussed here $pp \to pp \gamma$ process gives a sizeable
contribution to the low mass $(m_{X} > m_{p})$ single diffractive cross section.

\begin{figure}[!ht]
\includegraphics[width = 0.49\textwidth]{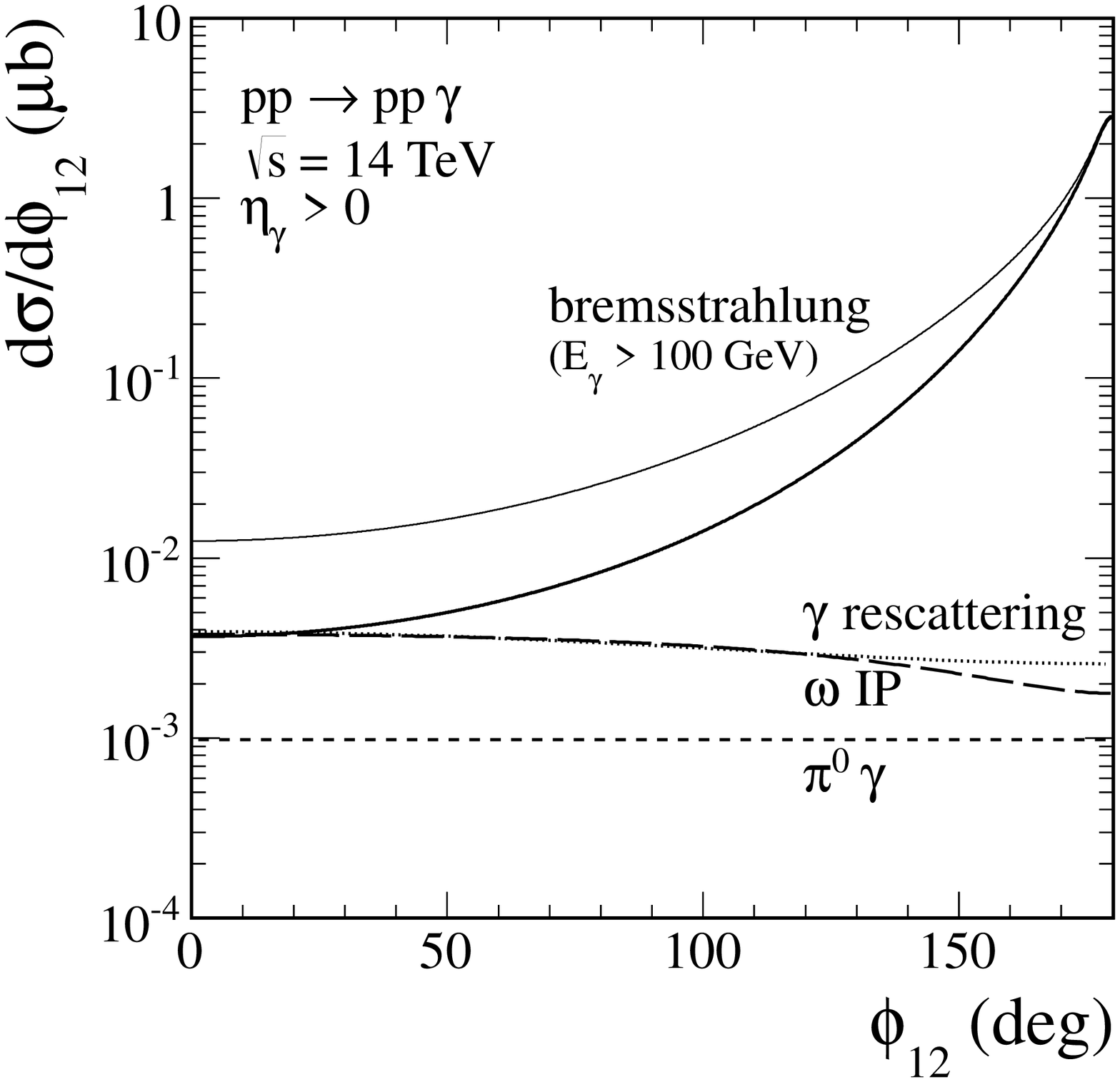}
  \caption{\label{fig:dsig_dphi}
  \small
Distribution in relative azimuthal angle between outgoing protons
for different mechanisms at $\sqrt{s} = 14$~TeV and $\eta_{\gamma}>0$.
For classical bremsstrahlung we have imposed $E_{\gamma} > 100$~GeV
and used two values of $\Lambda_{N} = 0.8, 1$~GeV of the proton off-shell form factors
(see the lower and upper solid line, respectively).
}
 \end{figure}

If both protons are measured one could also study correlations
in the relative azimuthal angle between outgoing protons. 
Our model calculations are shown in Fig.\ref{fig:dsig_dphi}. 
One can observe a large enhancement at back-to-back configurations 
for the classical bremsstrahlung which reminds the elastic scattering case 
($\phi_{12} = \pi$).
The contributions for other mechanisms are significantly smaller
and weakly depend on $\phi_{12}$.

\section{Conclusions}

In this paper we have considered several mechanisms of exclusive single photon production.
We have calculated several differential distributions 
for the $p p \to p p \gamma$ reaction at high energy for the first time in the literature.
The classical bremsstrahlung mechanism turned out to give the biggest
cross section concentrated at large photon (pseudo)rapidities. 
The photons are emitted at only slightly smaller pseudorapidities than the scattered protons.
We observe a strong cancellation between the initial and final state radiation.
The cross section for the classical bremsstrahlung is peaked at 
back-to-back configurations (similar transverse momenta or polar angles of 
outgoing protons and relative azimuthal angle concentrated close to
$\phi_{12} = \pi$).
This is a clear reminiscence of elastic scattering. 
Cut on photon energy ($E_{\gamma} > 100$~GeV) reduces 
the region of $\phi_{12} \cong \pi$ significantly
and the integrated diffractive bremsstrahlung cross section
is only of the order of $\mu$b.
The cross section for pion-photon or photon-pion exchanges is much smaller.
Here both small (photon exchange) and large (pion exchange)
four-momentum transfers squared are possible. 
For this process there is no correlation in azimuthal angle between outgoing protons.

Both classical bremsstrahlung and pion-photon (photon-pion)
as well as virtual-omega rescattering mechanisms
could be studied with the help of Zero Degree Calorimeters (photons)
and the ALFA or TOTEM detectors (protons).
By imposing several cuts one could select or enhance the contribution of one of the mechanisms. 
The cross section for pomeron-photon or photon-pomeron exchanges is rather small
and concentrated at midrapidities. 
Furthermore, the transverse momenta of outgoing 
photons are small and cannot be easily measured with central ATLAS or CMS detectors.

Summarizing, even present LHC equipment allows to study
exclusive production of photons. Since this process was never
studied at high energies it is worth to make efforts to obtain
first experimental cross sections. 
Since the cross sections are reasonably large 
one could try to obtain even some differential distributions.
This would allow to test our understanding of the diffractive processes 
and help in pinning down some hadronic and electromagnetic off-shell form factors, 
difficult to test otherwise.

\vspace{0.5cm}
{\bf Acknowledgments}

We are indebted to W. Sch\"afer for a discussion of some theoretical aspects 
of our calculation and J. L\"{a}ms\"{a}, M. Murray and R. Orava
for a discussion of some experimental aspects.
This work was partially supported by the Polish grant
No. PRO-2011/01/N/ST2/04116.


\end{document}